%% file: paper.tex
\pdfoutput=1
\documentclass[a4paper,12pt]{article}
\usepackage{ifpdf}
\ifpdf
\usepackage[pdftex]{graphicx}
\usepackage[usenames,dvipsnames]{color}
\else
\usepackage{graphicx}
\usepackage[usenames,dvips]{color}
\fi
\usepackage{amsmath}
\usepackage{amssymb}
\usepackage[colorlinks=true]{hyperref}   
\usepackage{caption}
\usepackage{paper}

\numberwithin{equation}{section}

\begin{document}

\title{A numerical study of the\\
Regge Calculus and Smooth Lattice methods\\
on a Kasner cosmology.}
\author{%
Leo Brewin\\[10pt]%
School of Mathematical Sciences\\%
Monash University, 3800\\%
Australia}
\date{04-Nov-2014}
\date{05-Nov-2014}
\date{19-Dec-2014}
\date{20-Dec-2014}
\date{11-Jan-2015}
\date{29-Jan-2015}
\reference{Preprint: arXiv:2015.xxxx}

\maketitle

\begin{abstract}
\noindent
Two lattice based methods for numerical relativity, the Regge Calculus and the Smooth
Lattice Relativity, will be compared with respect to accuracy and computational speed in a
full 3+1 evolution of initial data representing a standard Kasner cosmology. It will be
shown that both methods provide convergent approximations to the exact Kasner cosmology. It
will also be shown that the Regge Calculus is of the order of \CpuRatio\ times slower than
the Smooth Lattice method.
\end{abstract}

\section{Introduction}
\label{sec:Intro}

This is a good time to be doing numerical relativity. Most of the important hard problems
have largely been solved to the extent that computations with important astrophysical
applications are now treated, to a degree, as routine. But past experience with
computational physics indicates that new algorithms will be required and thus it seems
timely to revisit alternative approaches to numerical relativity.

Lattice based methods, such as the Regge Calculus \cite{regge:1961-01,gentle:2002-01}, have
most commonly been used as a possible basis for quantum gravity and, to a lesser extent, in
numerical relativity. They provide an elegant distinction between the topological properties
of the lattice (by way of the connections between the vertices) and the geometry of the
lattice (by the assignment of lengths to the legs). The ease with which complex topologies
can be encoded in a lattice is often cited as a clear advantage of lattice methods over
traditional grid based methods (though this claim has yet to be demonstrated in a
non-trivial example).

In this paper two lattice based methods, the Regge Calculus
\cite{regge:1961-01,gentle:2002-01,wheeler:1964-01} and Smooth Lattice Relativity
\cite{brewin:2010-02,brewin:2010-03,brewin:2009-05} will be compared head to head with
particular emphasis on the computational costs and to a lesser extent the accuracy of both
methods for a simple $T^3$ Kasner cosmology. Our results show (see section
(\ref{sec:Results})) that both methods provide convergent approximations to the continuum
but at vastly different computational cost -- the Regge Calculus is around \CpuRatio\ times
slower than the smooth lattice method. This is a severe limitation that precludes meaningful
comparison of the two methods for less symmetric space-times. Comparisons of the Smooth
Lattice Method with traditional finite difference methods for the Teukolsky, Brill and Gowdy
space-times will be presented elsewhere.

\section{Smooth Lattice Relativity}
\label{sec:SLGR}

Since the smooth lattice method is not well known it is reasonable to take a short moment to
describe its basic features.

Put simply, the smooth lattice is a discrete approximation to some possibly unknown smooth
geometry. In the case where the smooth geometry is known explicitly it is rather easy to
construct a discrete approximation, the smooth lattice, from the given smooth geometry. The
smooth geometry is required only to provide the necessary information, its topology and
metric, to allow the construction to proceed. It serves no real purpose after the
construction of the lattice (though it might reappear when questions of convergence are
addressed).

Consider some given smooth geometry composed of a smooth manifold equipped with a smooth
metric. How might a discrete approximation to that geometry be built? The short answer is
that the manifold can be approximated by a finite lattice and the metric by an assignment of
lengths to the legs of the lattice. But how is that lattice constructed? And how is the
assignment of leg lengths made?

The lattice can be built by drawing upon familiar ideas from differential geometry. Recall
that an atlas on a manifold consists of a sequence of overlapping charts and transition
functions between pairs of neighbouring charts. Given any atlas, a lattice can be
constructed simply by introducing one vertex per chart and connecting each vertex to its
nearest neighbours (these connections will be the legs of the lattice). Other choices are of
course possible (e.g., by adding more vertices in each chart) but this serves as a simple
example. Now consider the metric and its encoding on the lattice. It is tempting to declare
that each leg of the lattice is also a geodesic segment of the continuum geometry. But doing
so introduces a minor problem -- there may exist legs in the lattice which fail to be
described by a unique geodesic. Fortunately this is easily fixed by simply refining the
charts into smaller and smaller charts to the point that every leg in the lattice is
described by a unique geodesic. It is well known that it is always possible to do so (in the
absence of curvature singularities).

The final step in this construction is to adopt local Riemann normal coordinates in a
neighbourhood of a selected vertex. This is done for a number of reasons
\begin{itemize}
\item it captures the essence of the Einstein equivalence principle,
\item it guarantees that, in the absence of space-time singularities, the Riemann components
      are bounded,
\item it reduces covariant derivatives to partial derivatives and a consequent reduction in
      the complexity of many differential operators.
\end{itemize}

In the local Riemann normal coordinates, $x^\mu = (t,x,y,z)^\mu$, the metric can be written
as
\begin{align}
\gab(x) &= \gab
           - \frac{1}{3}\Racbd x^\mu x^\nu
           + \frac{1}{12}\dRacbde x^\mu x^\nu x^\gamma
           + \BigO{L^4}
\label{eqn:RNCgab}
\end{align}
where $L$ is a typical length scale for the neighbourhood of the vertex and
$\gab=\diag(-1,1,1,1)$. This choice of $\gab$ ensures that the coordinate basis vectors
$\partial_a$ form an orthonormal basis for the tangent space at the chosen vertex.

It is a straightforward computation to show, given the above form of the metric, that the
length $\Lij$ of the geodesic segment connecting vertices $i$ and $j$ is given by
\cite{brewin:2009-03,willmore:1996-01}
\begin{align}
\Lij^2 &= \gab \Dx^\alpha_{ij} \Dx^\beta_{ij}
        - \frac{1}{3}\Racbd x^\alpha_i x^\beta_j x^\mu_i x^\nu_j
        + \BigO{L^5}
\label{eqn:RNCLij}
\end{align}
These equations can be used to compute the coordinates of each vertex $x^\alpha_i$ given the
$\Lij$ and $\Rabcd$ as described in section (6) of \cite{brewin:2010-03}. The translational
and rotational symmetries are accounted for by locating the coordinate origin at the chosen
vertex while aligning selected coordinate axes with specific legs of the lattice.

The construction just given will produce a smooth lattice that is discrete in both space and
time. However, it is standard practice in numerical relativity to cast the Einstein
equations in the form of a Cauchy initial value problem. This imposes a small change in the
process described above. Prior to introducing the lattice, the smooth 4 dimensional
space-time is assumed to be foliated into a sequence of Cauchy surfaces. A 3 dimensional
lattice is built from a set of vertices in an initial Cauchy surface and subsequently
propagated onto future Cauchy surfaces by the Einstein equations and suitable gauge
conditions. In this picture the leg lengths, Riemann curvatures and coordinates are now
regarded as functions of time. Importantly the legs and the Riemann curvatures retain their
4 dimensional status (i.e., each leg is a geodesic of the 4-dimensional metric not the
3-metric of the Cauchy surface, each leg is a chord of the space-time connecting pairs of
vertices in the Cauchy surface). This is perhaps not obvious at this point but will be made
clear later after introducing the full set of evolution equations.

In making the transition from concepts based in differential geometry to a discrete lattice
it helps to introduce some new terminology and notation (to emphasise the distinction between
continuum and discrete structures). Thus a neighbourhood of a selected vertex will be known
as a cell on a central vertex. The cell will consist of its central vertex together with its
immediate neighbouring vertices and the set of legs shared by those vertices. A frame will
be defined as a cell together with a set of geometrical data for that cell including (no
less than) a local set of coordinates, the leg lengths for each leg in the cell and the
Riemann curvatures at the central vertex.

Cells will overlap and this requires some care when specifying frame dependent quantities
(such as tensor components). The following notation will be introduced to avoid any
ambiguity. A quantity $R$, defined on vertex $q$ in the frame of vertex $p$, will be denoted
by $R_{q\barp}$. The subscripts $q\barp$ will be dropped in cases where no ambiguity can
arise.

Vertices in a cell will be denoted by lowercase Latin letters while Greek letters will be
used for all tensor indices.

\subsection{The SLGR evolution equations}

In an earlier paper \cite{brewin:2010-03} a set of evolution equations for the lattice were
proposed in which the dynamical variables were the leg lengths, their first time derivatives
and the Riemann curvatures. The extrinsic curvatures and vertex coordinates were treated as
kinematical variables and were computed as solutions of simple algebraic equations (see
sections (6.1) and (7.1) of \cite{brewin:2010-03}). The experience gained since then shows
that there are better choices for the dynamical variables leading to greatly simplified
evolution equations and considerably reduced computational cost. Two choices for the
dynamical variables will be presented. The first uses $(\Lij,\Kab,\Rabcd)$ as dynamical
variables while the second uses $(x^\mu_i,\Kab,\Rabcd)$. In both cases equations
(\ref{eqn:RNCLij}) are used to compute the remaining data, either $x^\mu_i$ or $\Lij$, from
the dynamical variables.

In the case of a unit lapse and zero shift vector, as used throughout this paper, the
evolution equations for the leg lengths and extrinsic curvatures (see equations (3.1,3.2) of
\cite{brewin:2010-03}) are
\begin{align}
	\frac{d\Lij^2}{dt} &= -2 \Kab\Dx^\alpha_{ij}\Dx^\beta_{ij}\label{eqn:SLGRdotLij}\\[5pt]
	\frac{d}{dt}\left(\Kab\Dx^\alpha_{ij}\Dx^\beta_{ij}\right)
	   &= \left(-K_{\mu\alpha}K^\mu{}_{\beta}
	            +R_{\alpha t\beta t}\right)\Dx^\alpha_{ij}\Dx^\beta_{ij}\label{eqn:SLGRdotKij}
\end{align}
where $\Dx^\alpha_{ij}=x^\alpha_i - x^\alpha_j$, $R_{\mu t\nu
t}=R_{\mu\alpha\nu\beta}n^\alpha n^\beta$ and $n^\mu=\delta^\mu{}_t$ is the future pointing
unit time-like normal to the Cauchy surface at the central vertex.

The normal use of equations (\ref{eqn:SLGRdotLij},\ref{eqn:SLGRdotKij}) would be to dictate
the evolution of legs in the lattice. There is, however, no reason why that pair of
equations can not be applied to any leg whether or not it happens to be defined by a pair of
vertices of the lattice. This simple observation can be put to good use to obtain explicit
evolution equations for each of the extrinsic curvatures. As an example, consider a
fictitious leg defined by the coordinates $(0,0,0,0)$ and $(0,\Lxx,0,0)$. When substituted
into (\ref{eqn:SLGRdotLij},\ref{eqn:SLGRdotKij}) this leads to the following pair of
evolution equations
\begin{align}
	\frac{d\Lxx^2}{dt} &= -2 \Kxx \Lxx^2\\[5pt]
	\frac{d}{dt}\left(\Kxx\Lxx^2\right) &= -K_{x\alpha}K^{\alpha}{}_{x}\Lxx^2 + \Rtxtx\Lxx^2
\end{align}
which, upon eliminating $\Lxx$, leads to
\begin{align}
	\dotKxx = \Kxx^2 - \Kxy^2 - \Kxz^2 + \Rtxtx
\label{eqn:SLGRdotKij1}
\end{align}
This same idea can be employed for the remaining extrinsic curvatures with the result that
\begin{align}
	\dotKyy &= \Kyy^2 - \Kxy^2 - \Kyz^2 + \Rtyty\label{eqn:SLGRdotKij2}\\[5pt]
	\dotKzz &= \Kzz^2 - \Kxz^2 - \Kyz^2 + \Rtztz\label{eqn:SLGRdotKij3}\\[5pt]
	\dotKxy &= -\Kxz\Kyz + \Rtxty\label{eqn:SLGRdotKij4}\\[5pt]
	\dotKxz &= -\Kxy\Kyz + \Rtxtz\label{eqn:SLGRdotKij5}\\[5pt]
	\dotKyz &= -\Kxy\Kxz + \Rtytz\label{eqn:SLGRdotKij6}
\end{align}
(for the mixed terms such as $d\Kxy/dt$, use a fictitious leg joining $(0,0,0,0)$ and
$(0,\Lxx,\Lyy,0)$).

The evolution equations for the coordinates can be obtained by basic arguments (see Appendix
\ref{app:EvolveRNCs} for full details). The result, for the vertex $q$ in the frame of
$\barp$, is
\begin{align}
   \frac{dx^\mu_{q\bar p}}{dt} = -K^\mu{}_\nu x^\nu_{q\bar p} + \BigO{L^2}\qquad\hbox{for }\mu=x,y,z
\label{eqn:SLGRdotXcoord}
\end{align}

The final set of evolution equations for the lattice are those for the Riemann
curvatures. These can be obtained from equations (4.4) to (4.17) of \cite{brewin:2010-03}
which, in the simple case of a unit lapse function, are given by
\bgroup
\def\P{\phantom{+}}
\def\M{-}
\begin{align}
\dotRxyxy &=  \P \dRtyxyx - \dRtxxyy&
\dotRxyxz &=  \P \dRtzxyx - \dRtxxyz
\label{eqn:SLGRdotRabcd1}\\[5pt]
\dotRxyyz &=  \P \dRtzxyy - \dRtyxyz&
\dotRxzxz &=  \P \dRtzxzx - \dRtxxzz
\label{eqn:SLGRdotRabcd2}\\[5pt]
\dotRxzyz &=  \P \dRtzxzy - \dRtyxzz&
\dotRyzyz &=  \P \dRtzyzy - \dRtyyzz
\label{eqn:SLGRdotRabcd3}\\[5pt]
\dotRtxxy &=  \M \dRxyxyy - \dRxyxzz&
\dotRtyxy &=  \P \dRxyxyx - \dRxyyzz
\label{eqn:SLGRdotRabcd4}\\[5pt]
\dotRtzxy &=  \P \dRxyxzx + \dRxyyzy&
\dotRtxxz &=  \M \dRxyxzy - \dRxzxzz
\label{eqn:SLGRdotRabcd5}\\[5pt]
\dotRtyxz &=  \P \dRxyxzx - \dRxzyzz&
\dotRtzxz &=  \P \dRxzxzx + \dRxzyzy
\label{eqn:SLGRdotRabcd6}\\[5pt]
\dotRtyyz &=  \P \dRxyyzx - \dRyzyzz&
\dotRtzyz &=  \P \dRxzyzx + \dRyzyzy
\label{eqn:SLGRdotRabcd7}
\end{align}
\egroup
The above equations (\ref{eqn:SLGRdotRabcd1}--\ref{eqn:SLGRdotRabcd7}) are nothing more than
the second Bianchi identities coupled with the vacuum Einstein equations. The use of partial
derivatives rather than covariant derivatives stems from the use of Riemann normal
coordinates (in which the connection vanishes).

In summary, the evolution equations for the first set of dynamical variables
$(\Lij,\Kab,\Rabcd)$ are (\ref{eqn:SLGRdotLij},\ref{eqn:SLGRdotKij1}--\ref{eqn:SLGRdotKij6})
and (\ref{eqn:SLGRdotRabcd1}--\ref{eqn:SLGRdotRabcd7}) while for the second set of dynamical
variables, $(x^\mu_i,\Kab,\Rabcd)$, the evolution equations are
(\ref{eqn:SLGRdotXcoord},\ref{eqn:SLGRdotKij1}--\ref{eqn:SLGRdotKij6}) and
(\ref{eqn:SLGRdotRabcd1}--\ref{eqn:SLGRdotRabcd7}).

In the following these two sets of evolution equations will be referred to as evolution
schemes 1 and 2 respectively.

\subsection{The SLGR source terms}
\label{sec:SLGRSourceTerms}

Given that the Riemann curvature components are known at each vertex of the lattice it would
seem that the source terms in (\ref{eqn:SLGRdotRabcd1}--\ref{eqn:SLGRdotRabcd7}) could be
easily evaluated using a suitable finite difference scheme. There is however one important
issue that must be noted. In any cell the point values of the Riemann curvatures are known
only at the central vertex. The values on the remaining vertices are with respect to the
local frames associated with neighbouring cells. Thus before the finite differences are
taken, the curvatures must first be imported, from the neighbouring frames, into the frame
of the chosen cell. Fortunately this is rather straightforward task. The key observation is
that a pair of neighbouring cells will share a set of legs. Choose one of the vertices and
pick four of the shared legs attached to that vertex. To each leg construct a tangent vector
and its corresponding components with respect to each frame. Assuming that this set of
vectors are linearly independent (it will argued later in section (\ref{app:SourceTerms})
that this assumption will almost always be satisfied) then there exists a unique map between
the two frames (at this vertex). This map between the frames is the discrete analog of the
transition functions from the continuum. It is this map that is used when importing data
from one frame to another. See section (\ref{app:BiCubic}) for full details on how this map
can be computed for the bi-cubic lattice.

For any chosen cell this procedure will produce, at each vertex of the cell, the components
of the Riemann curvature in the frame of that cell. This data can then be used to estimate
the partial derivatives at the central vertex. Note that the data will, in general, not lie
on a regular grid thus the best that can be hoped for is for first order accurate estimates
in the derivatives (i.e., an $\BigO{L}$ truncation error). This is not ideal but is the best
that can be obtained with nearest neighbour interactions.

\section{The Regge calculus}
\label{sec:ReggeCalc}

The Regge Calculus and the Smooth Lattice method are built on a common structure -- they
both use a lattice and a table of leg lengths in forming a discrete approximation to a
continuum geometry. The principal difference between the two approaches lies in the nature
of the metric assigned to the lattice. The Regge calculus requires that the metric be
piecewise flat while the Smooth Lattice methods uses a locally flat approximation. The
curvature in a piecewise flat metric must be treated as a distribution with support on the
two dimensional subspaces of the lattice (commonly known as bones or hinges and are usually
the triangular faces of the 4-simplices of the lattice; the 4-simplex is the canonical cell
in a Regge lattice). Working with distribution valued quantities in a non-linear theory such
as General Relativity is a mathematically delicate area and requires considerably care. The
upshot is that the Einstein equations can not be easily imposed onto the Regge lattice.
However it is possible \cite{cheeger:1984-01} to unambiguously evaluate the Hilbert action
on a Regge lattice, leading to
\begin{align}
	I = \sum_{M}\>\theta A
\label{eqn:ReggeAction}
\end{align}
where the sum includes all of the bones within the lattice $M$ and $\theta$ and $A$ are the
defect angle and area of a typical bone (both of which can be computed from the known leg
lengths). Then, in analogy with the continuum case, the evolution equations for the lattice,
the Regge equations, are normally obtained by extremising this action with respect to the
leg lengths. Extremising with respect to leg $\Lij$ leads to
\begin{align}
	0 = \sum_{M(\Lij)}\>\theta\frac{\partial A}{\partial \Lij}
\label{eqn:ReggeEqtn}
\end{align}
where the sum now includes just the bones attached to the leg. There is one such equation
for each leg of the lattice. See \cite{regge:1961-01} for full details.

\subsection{The Regge evolution equations}

The equations just given are the full set of evolution equations for the Regge Calculus.
Though the equations are elegant they do present three hurdles. The first is one of
computational complexity -- the defects as functions of the legs are so involved that it is
simply not possible to present explicit expressions for the defects in terms of the leg
lengths (though the full details of the algorithm can be found in \cite{brewin:2010-01}).
The second hurdle concerns the way in which the Regge equations are solved -- they are a
fully coupled non-linear set of algebraic equations for the 4-dimensional leg lengths.
Gentle and Miller \cite{gentle:1998-01} employ a Sorkin evolution scheme
\cite{sorkin:1975-01,arXiv:gr-qc/9411008v1} in which a pair of existing Cauchy surface are
used to push forward to a future Cauchy surface. The Sorkin scheme provides a very elegant
means of solving the Regge equations though it does require some careful bookkeeping. The
final hurdle is more conceptual -- how can account be taken of the freedom to choose the
lapse and shift vector? This point is addressed in detail by Gentle and Miller
\cite{gentle:1998-01} where they argue that as the continuum limit is approached the above
Regge equations must reveal four degrees of freedom at each vertex and thus four equations
at each vertex must become redundant (in the continuum limit). They identify the four
equations and provide details of how those equations can be used as part of the evolution
scheme (in effect these equations propagate the vertices of one Cauchy surface while the
remaining Regge equations propagate the leg lengths). The Sorkin evolution scheme, as
implemented in this paper, is identical to that used by Gentle and Miller but with two minor
exceptions. Firstly, in this paper the continuum metric is used to set the initial data (the
consequences that follow from this choice will be discussed later in section
(\ref{sec:Results})). Secondly, where Gentle and Miller identify three pairs of equations
for the shift equations while later discarding one equation in each pair the approach taken
in this paper is to take the average of each pair.

\section{Initial data}
\label{sec:InitialData}

It was previously noted that topological and geometrical properties of a lattice can be
cleanly separated. This allows the initial data to be constructed in a two stage process.
First, choose a lattice with the required topology. Then add to that (raw) lattice the
required geometric data such as leg-lengths and, for the smooth lattice method, the Riemann
and extrinsic curvatures.

\subsection{The lattice}

\begin{figure}[ht]
\hbox to\textwidth{%
\hfill%
\includegraphics[width=0.45\textwidth]{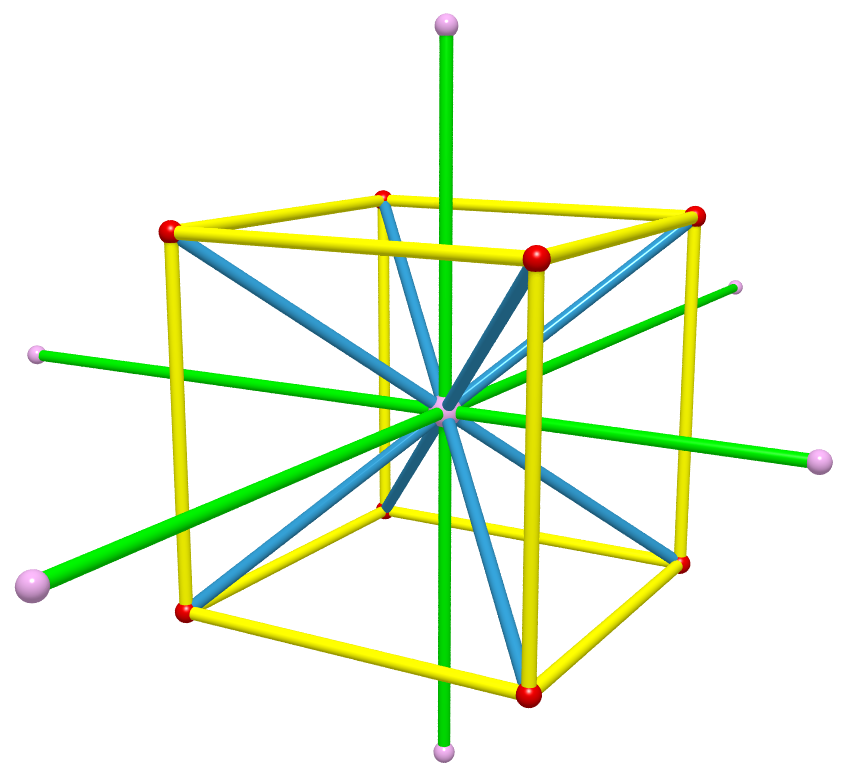}%
\hfill%
\raise20pt\hbox to 0.45\textwidth{\hfill\includegraphics[width=0.35\textwidth]{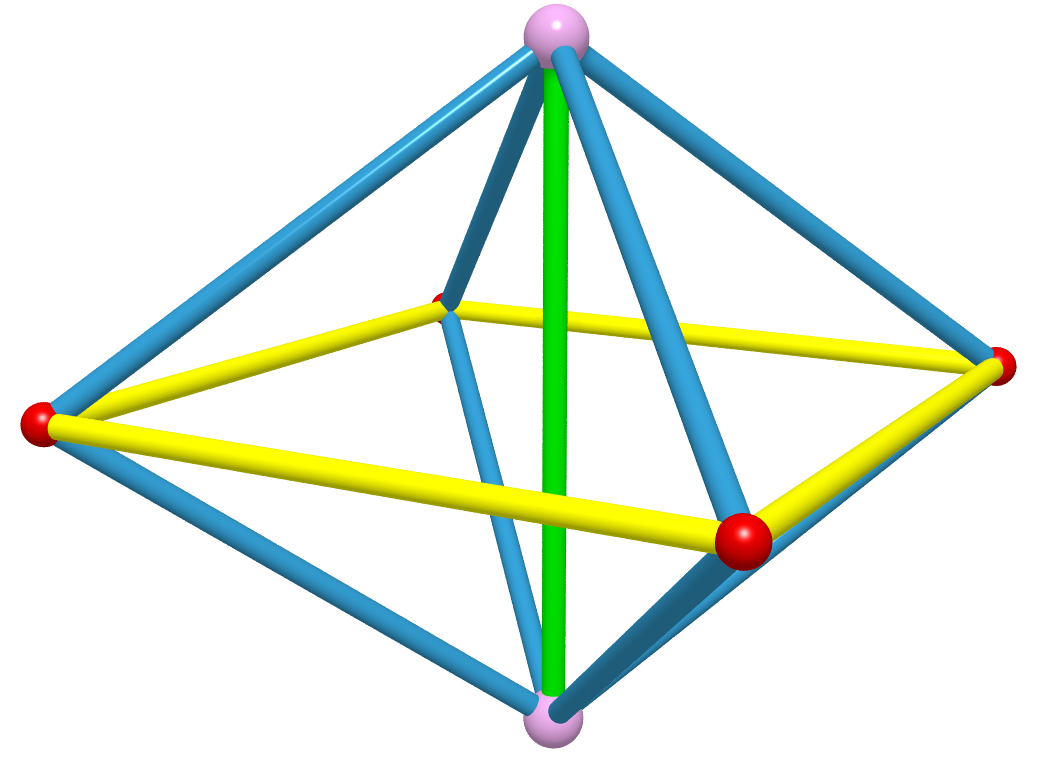}\hfill}%
\hfill%
}
\caption{\normalfont%
The local structure of the bi-cubic lattice. The lattice consists of two sub-lattices, one
defined by the purple vertices, the other by the red vertices. These are the $A$ and $B$
lattices described in the text. The left figure displays the legs associated with one cell.
The right figure shows the legs shared by a pair of cells.}
\label{fig:BiCubicLattice}
\end{figure}

Following Gentle and Miller \cite{gentle:1998-01}, each Cauchy surface will be modelled by a
bi-cubic lattice with opposite faces identified (as required by the $T^3$ topology). This
lattice is well suited to this cosmology as it allows not only to create initial data that
are manifestly homogenous but also to create a family of arbitrarily refined lattices (so
that convergence properties can be easily studied). The local structure of the lattice is
shown in figure (\ref{fig:BiCubicLattice}).

There is one feature of this lattice that requires special mention. The set of cells can be
split into two distinct groups. An element in one group is connected to another element in
the same group by following legs aligned to the cube (e.g., a horizontal leg in figure
(\ref{fig:BiCubicLattice})). While crossing from one group to the other requires moving
along a diagonal leg. In principle this distinction could be ignored but in practice there
is one advantage to be had. Denote the two groups by $A$ and $B$. The original lattice is
the sum of this pair. It is possible to contemplate using the smooth lattice method on just
group $A$ or on the pair $A$ and $B$. In each case the discretisation scale is the same but
the later case would require twice the work (for no likely gain in accuracy). For this
reason it was decided to apply the smooth lattice method to just one group (the choice is
unimportant).

A little more work is needed for the Regge calculus as it requires a fully 4-dimensional
lattice, bounded by two Cauchy surfaces and fully sub-divided into 4-simplices (a form of a
thin-sandwich approach). This structure can be obtained by lifting, in stages, groups of
vertices from one Cauchy surface forward in time to the second Cauchy surface. The full
details can be found in \cite{gentle:1998-01}.

\subsection{Geometric data}
\label{sec:GeomData}

Here a slightly different approach is taken compared to that taken by Gentle and Miller.
They chose to set the initial data for their Regge lattice by solving the Regge initial data
equations (a Regge form of the Hamiltonian and momentum constraints). This is a non-trivial
task and requires the solution of a large system of non-linear algebraic equations. In this
paper a much simpler approach was taken by setting the initial data for both the Regge
calculus and the smooth lattice method directly from the exact form of the Kasner metric. It
should be emphasised that this is not strictly correct (as the constraint equations are not
necessarily satisfied) but as the computational cost of an evolution does not depend on how
the initial data was set this short cut should have no impact on whatever conclusions might
be drawn from the results (this point will be discussed further in section
(\ref{sec:Discuss})).

The particular form of the vacuum Kasner $T^3$ cosmology used in this paper is described by
\begin{align}
ds^2 = -d\ut^2 + \ut^{2a} d\ux^2 + \ut^{2b} d\uy^2 + \ut^{2c} d\uz^2
\label{eqn:KasnerMetric}
\end{align}
where $a=b=4/3$ and $c=-1/3$. The non-standard notation for the coordinates
$(\ut,\ux,\uy,\uz)$ was chosen simply to avoid confusion with the local Riemann normal
coordinates $(t,x,y,z)$ used in the smooth lattice equations. However, since both the local
Riemann and global frames use a unit lapse and zero shift vector there should be no
confusion in setting $t=\ut$.

Given the form of the Kasner metric (\ref{eqn:KasnerMetric}) it is a simple matter to
compute, for $t>0$, the various quantities employed in the lattice. The conversion to
Riemann normal form (for the curvatures) is best done by projection onto a local orthonormal
frame (i.e., onto the unit vectors parallel to the coordinate axes). This leads to
\begin{gather}
\Kxx = -at^{-1}\>,\quad
\Kyy = -bt^{-1}\>,\quad
\Kzz = -ct^{-1}
\label{eqn:RNCKasner1}\\[10pt]
\Rxyxy = abt^{-2}\>,\quad
\Rxzxz = act^{-2}\>,\quad
\Ryzyz = bct^{-2}
\label{eqn:RNCKasner2}
\end{gather}
The remaining components are either zero (e.g., $\Kxy=0$) or can be obtained from the above
using known symmetries (e.g., $\Ryxxy=-\Rxyxy$).

The initial 3 dimensional lattice was constructed as a cube of vertices evenly spaced in the
Kasner $(\ut,\ux,\uy,\uz)$ coordinates. Each vertex was assigned coordinates in the form
$(a\dx,b\dy,c\dz)$ with $(a,b,c)$ a set of integers subject to $0\le a\le N_x$, $0\le b\le
N_y$ and $0\le c\le N_z$. The integers $N_x$, $N_y$ and $N_z$ count the number vertices
along the respective coordinate axes while the $(\dx,\dy,\dz)$ are the coordinates
increments between neighbouring vertices. The $T^3$ topology was obtained by identifying
vertices on opposite faces of the cube, for example, by identifying $(0,b,c)$ with
$(N_x,b,c)$. This produces a cube of dimensions $(N_x\dx)\times(N_y\dy)\times(N_z\dz)$ in
the Kasner coordinate space. The leg-lengths on this cube were set by solving the two-point
boundary value problem for the geodesic equation while the Riemann and extrinsic curvatures
were set using the exact data given by equations
(\ref{eqn:RNCKasner1},\ref{eqn:RNCKasner2}). All of the initial data were set at $t=1$.

The evolution of the initial data described in this paper uses a zero shift vector and a
unit lapse. Thus the Kasner coordinates of a typical vertex will be of the form
$(1+p\dt,a\dx,b\dy,c\dz)$ where $p$ is a positive integer and $\dt$ is the time step between
successive Cauchy surfaces. These coordinates will be used to compute exact values of the
lattice data (leg-lengths, curvatures etc.) for comparison with the numerical evolutions.

\section{Evolution}
\label{sec:Evolution}

The Regge data was evolved following the method given by Gentle and Miller with the small
exception previously noted (where the average of the pair of shift equations were taken
rather than using just one equation). The smooth lattice equations were integrated using a
4th order Runge-Kutta scheme with a variable time step as described below
(\ref{sec:Results}).

For the smooth lattice method there remains one small issue -- how can a unique time
derivative be computed for legs that are shared by neighbouring cells? The simple answer is
to take an average over all of the contributing cells. Note also in equation
(\ref{eqn:SLGRdotLij}) the time derivative uses an off-centre estimate for the extrinsic
curvature. This can be improved by constructing a linear Taylor series to estimate the
$\Kab$ at the centre of the leg with the first spatial derivatives of the $\Kab$ computed in
exactly the same manner as those for the $\Rabcd$. Since the Kasner geometry is homogenous
this step was not expected to make any significant changes to the evolution of the lattice.

\section{Results}
\label{sec:Results}

A number of simple tests were conducted to verify that both methods gave the expected
results and to measure their rates of convergence to the continuum. In the first test the
codes were run over $1<t<11$ with the initial data set at $t=1$ using $L=0.005$ while the
evolutions used a variable time step $\dt=\min(\Lxx,\Lyy,\Lzz)/4$. Convergence tests were
performed by multiple integrations over $1<t<8$ each with successively smaller $L$ and
corresponding $\dt$. In particular, seven integrations were performed with a fixed time step
$\dt=L/5$ where $L=0.5/2^q$ and $q=1,2,3\cdots 7$. A fixed time step was chosen simply to
make it easier to stop the evolution at exactly $t=8$.

There are three sets of data to be discussed, one for the Regge calculus and two for the
smooth lattice method (one for each of the evolution schemes). For the moment the discussion
will focus on the Regge calculus results versus those of the first evolution scheme for the
smooth lattice method. The results for the second scheme will be presented at the end of
this section.

On the $N_x=N_y=N_z=8$ lattice used by Gentle and Miller both the Regge calculus and smooth
lattice methods produced extremely homogenous evolutions over $1<t<11$ with variations in
$\Lxx$ across the lattice of the order of the machine precision (which in this case was
$10^{-18}$). Similar behaviour was noted in the extrinsic and Riemann curvatures in the
smooth lattice results (no such data is readily available for the Regge calculus). The
$N_x=N_y=N_z=8$ lattice is a rather coarse lattice so the calculations were repeated on a
$N_x=512,N_y=N_z=8$ lattice with the results again being of the order of the machine
precision.

The evolution of the fractional error in $\Lxx$ and $\Lzz$ are shown in figure
(\ref{fig:ErrorsLegs}). In this and following figures, the fractional error $E(Q)$ in some
quantity $Q$ is defined by $E(Q) = 1 - Q/Q^e$ where the superscript $e$ denotes the exact
value (as computed from the Kasner metric). There is little to note here other than that the
errors are small and grow smoothly with time.

\begin{figure}[ht]
\FigPair{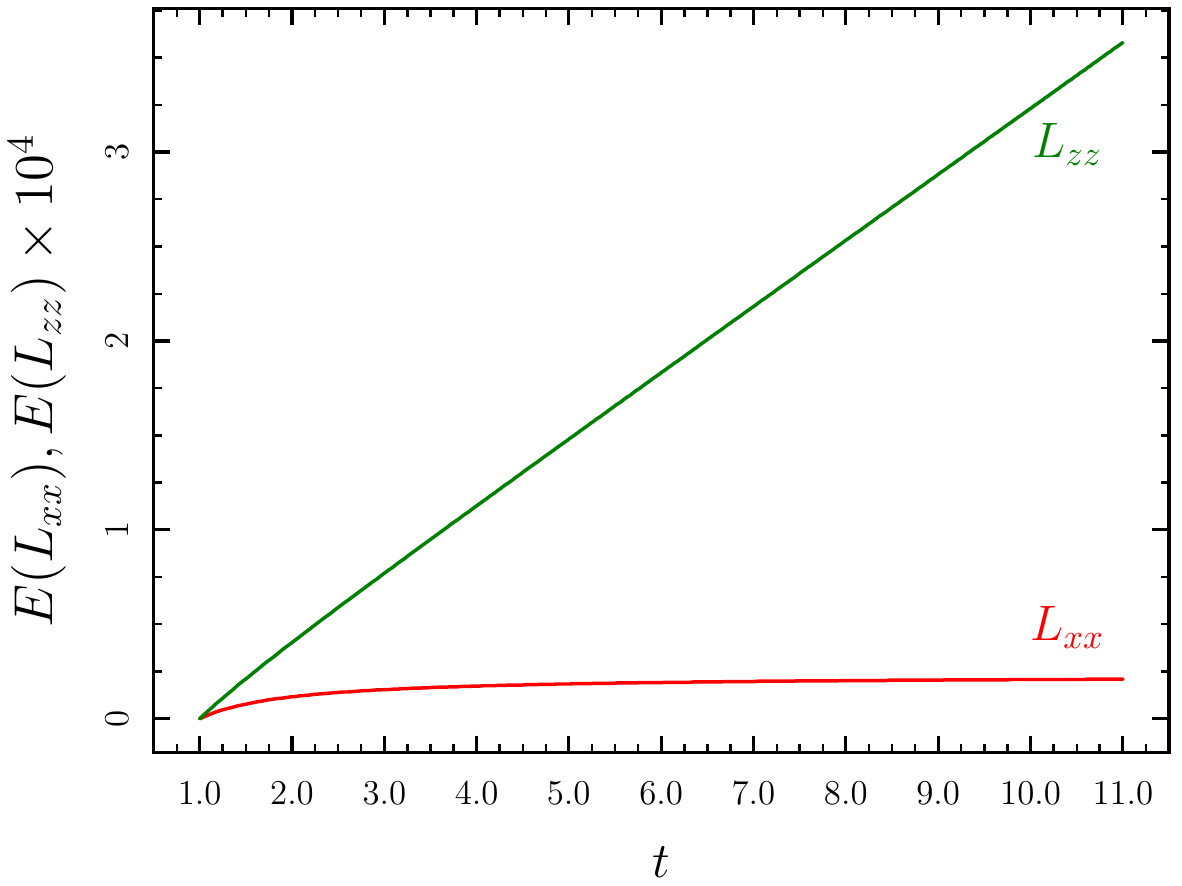}{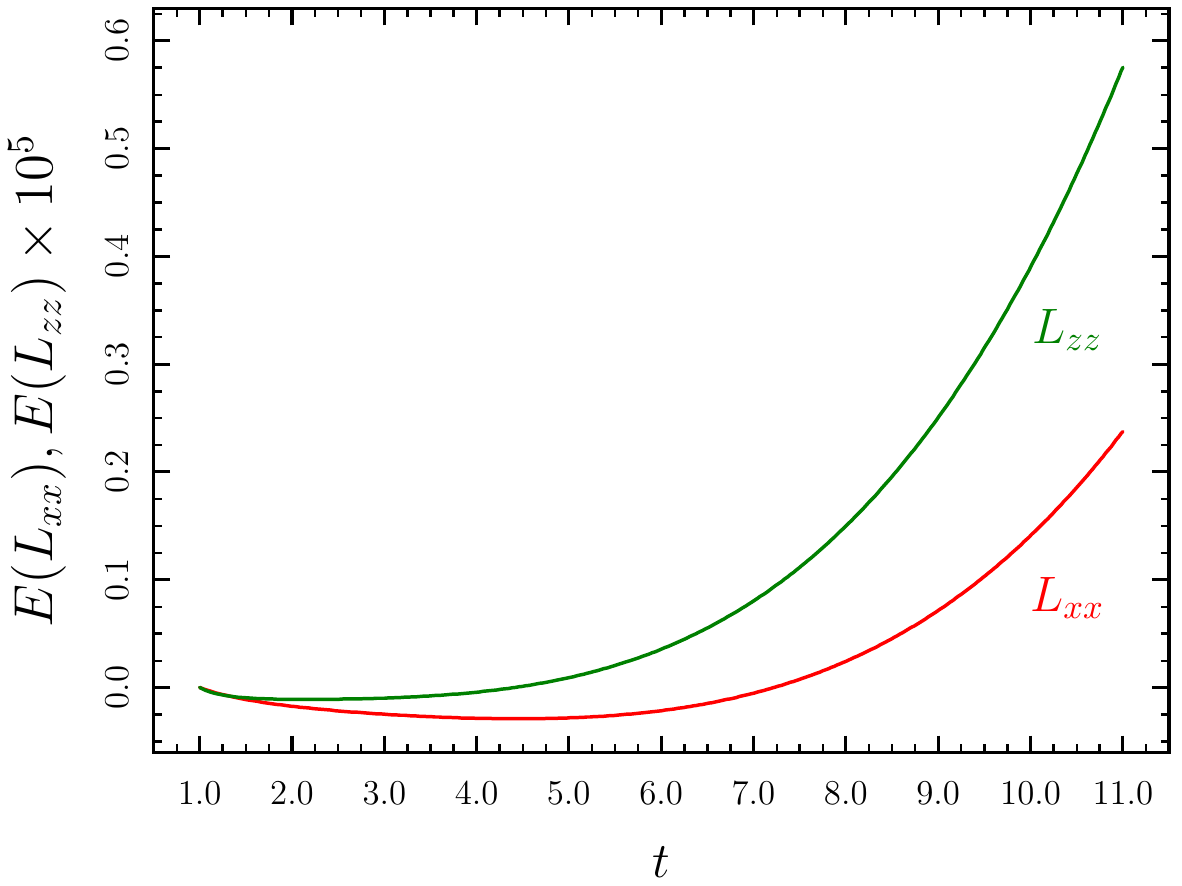}
\caption{\normalfont%
The fractional errors in the leg lengths for the Regge calculus (left) and
the smooth lattice method (right).}
\label{fig:ErrorsLegs}
\end{figure}

\begin{figure}[ht]
\FigPair{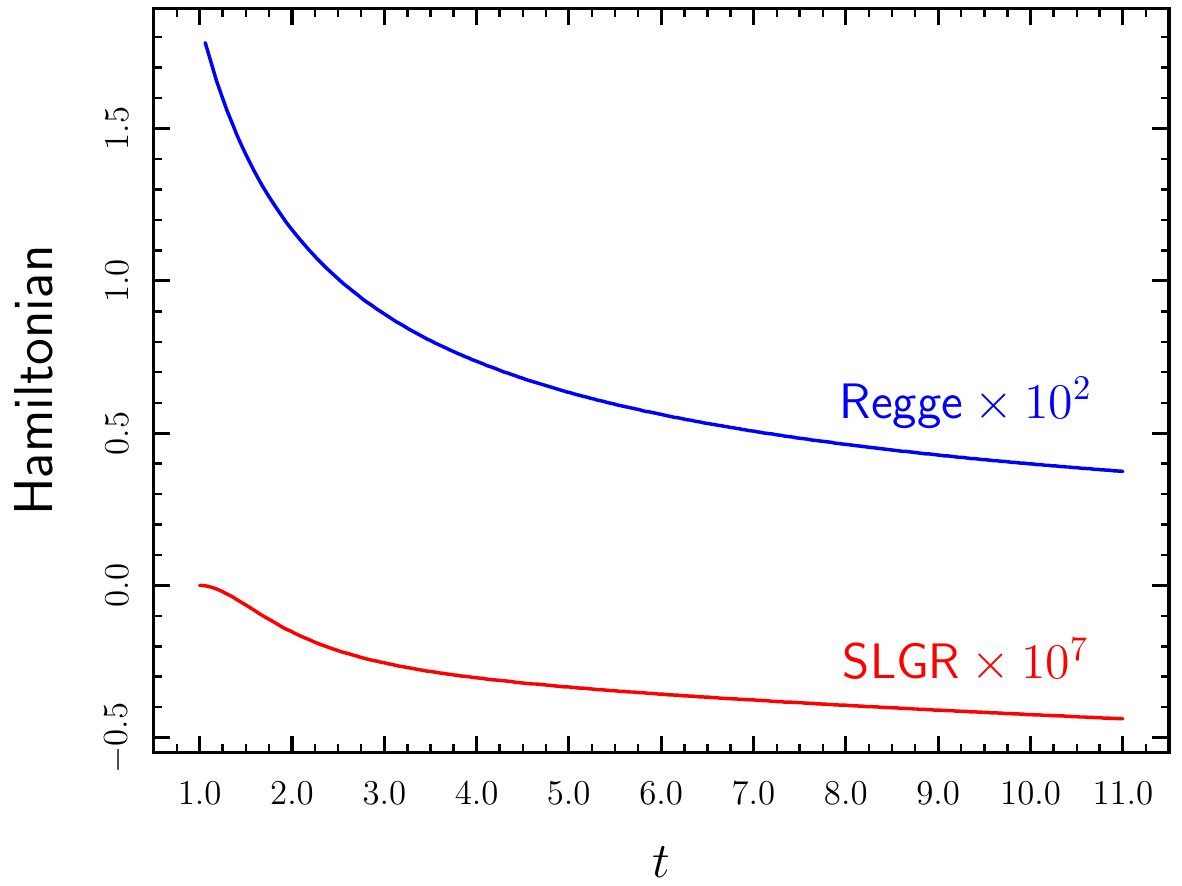}{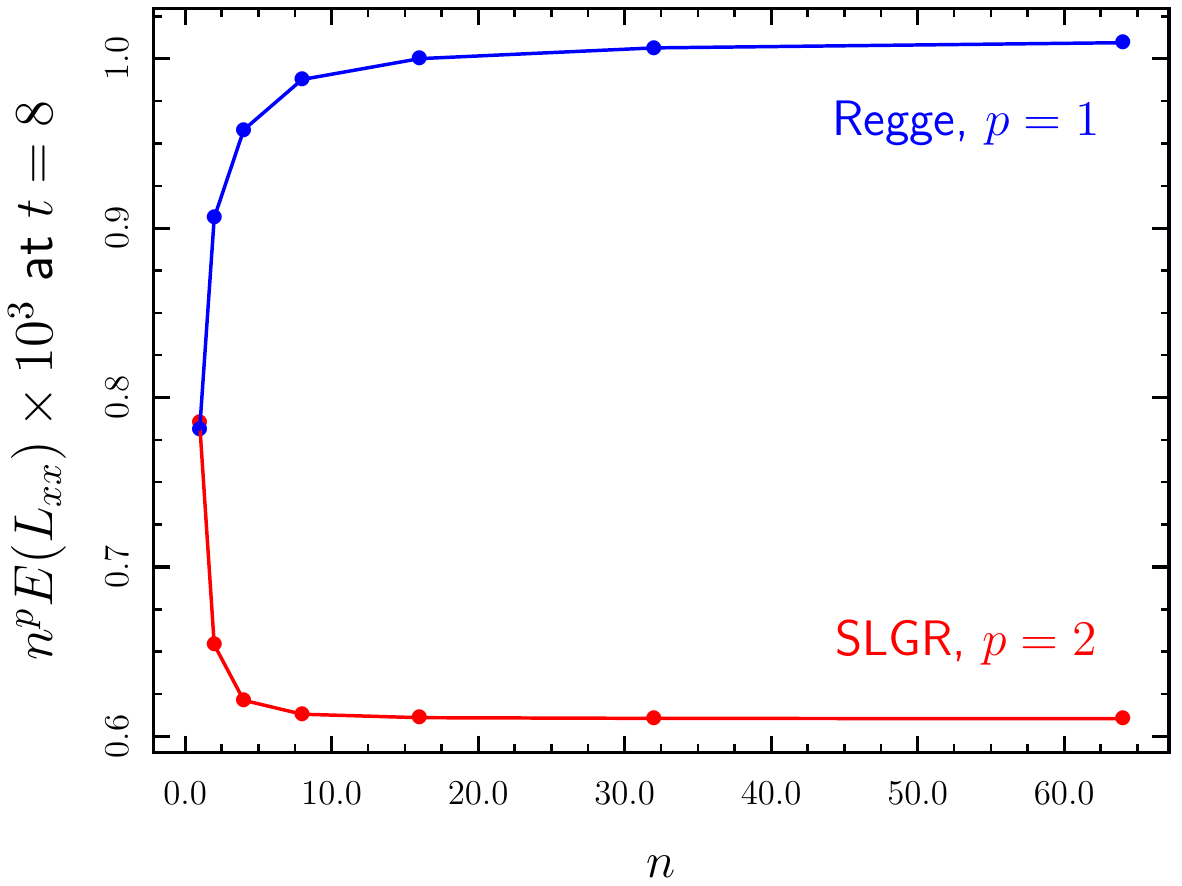}
\caption{\normalfont%
The Hamiltonian constraint (left) and the convergence estimate for the errors
in $\Lxx$ (right).}
\label{fig:HamiltonConverge}
\end{figure}

The Hamiltonian constraint is shown in figure (\ref{fig:HamiltonConverge}). The Regge
Hamiltonian, as described in detail by Gentle and Miller, was scaled by an estimate of the
volume per vertex $V=\Lxx\Lyy\Lzz$ (in the spirit of Wheeler's estimate of the 3-Riemann
scalar (\cite{wheeler:1964-01,piran:1986-01})). The Smooth Lattice Hamiltonian is taken to
be $H = 2(\Rxyxy + \Rxzxz + \Ryzyz)$. This differs from the more familiar form of the
Hamiltonian, namely $H = {}^{(3)}R+K^2 - K_{ab} K^{ab}$, for the simple fact that the smooth
lattice method works directly with the 4-Riemann curvatures rather than the 3-Riemann
curvatures. Figure (\ref{fig:HamiltonConverge}) shows that the initial value of the Regge
Hamiltonian is not zero. This is a direct consequence of setting the initial data via the
exact solution rather than by enforcing the Regge constraints.

The convergence of the lattice solution to the continuum is displayed in the right hand
panel of figure (\ref{fig:HamiltonConverge}) and it would appear that while the smooth
lattice method displays second order convergence the Regge calculus appears to be first
order convergent. This conflicts with the second order convergence reported by Gentle and
Miller. The most plausible explanation is that as the Regge initial data was not set by
enforcing the Regge constraints, a first order error in the initial data has been introduced
and that error has been propagated forward in time.

\begin{figure}[ht]
\FigPair{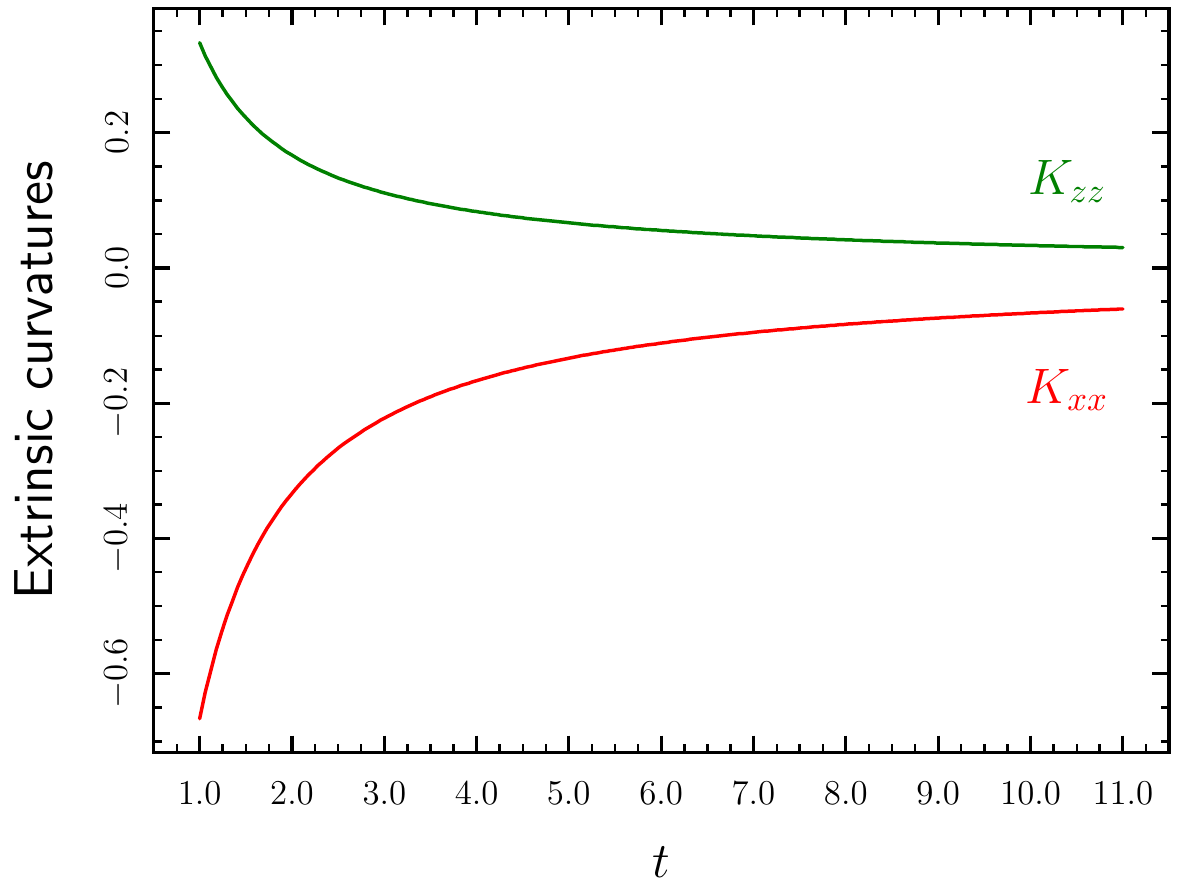}{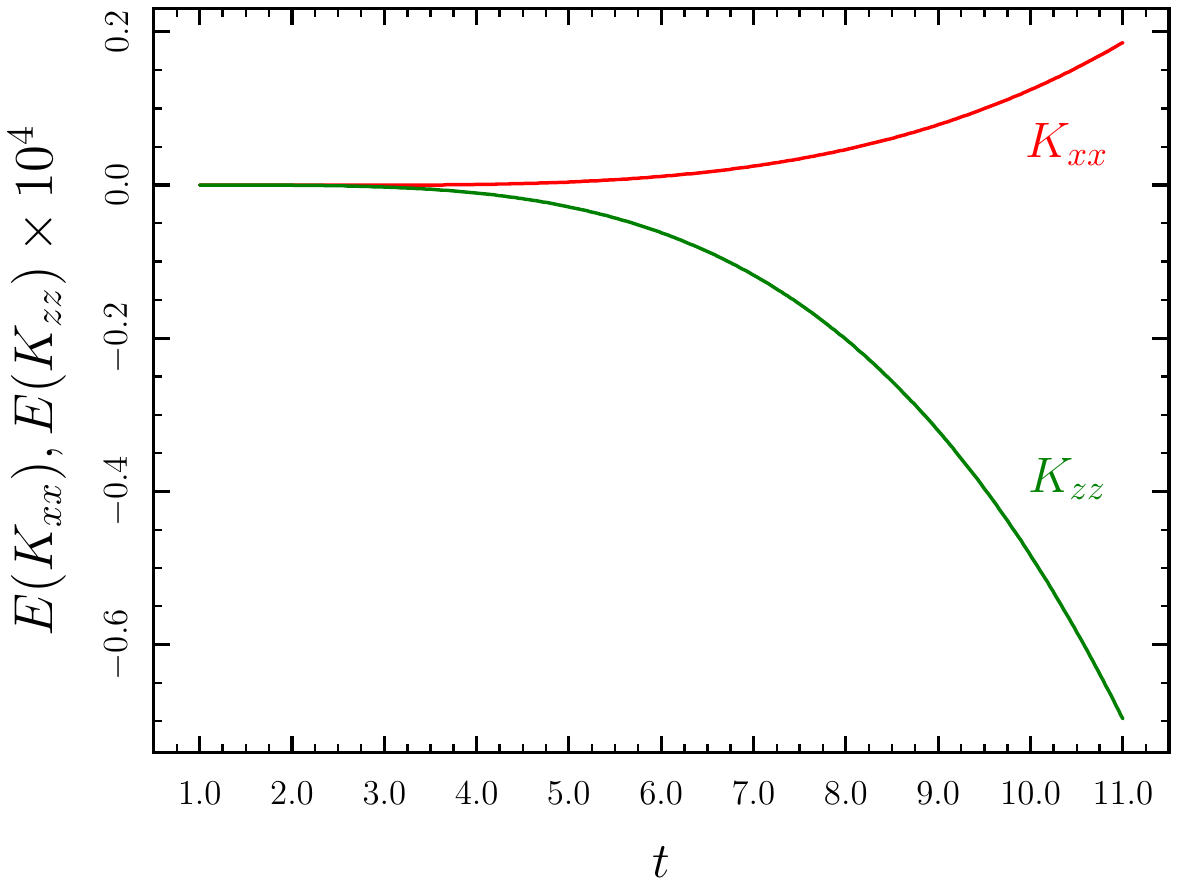}
\caption{\normalfont%
Smooth lattice evolution of $\Kxx$ and $\Kyy$ (left) and their corresponding
fractional errors (right).}
\label{fig:slgrKab}
\end{figure}

\begin{figure}[ht]
\FigPair{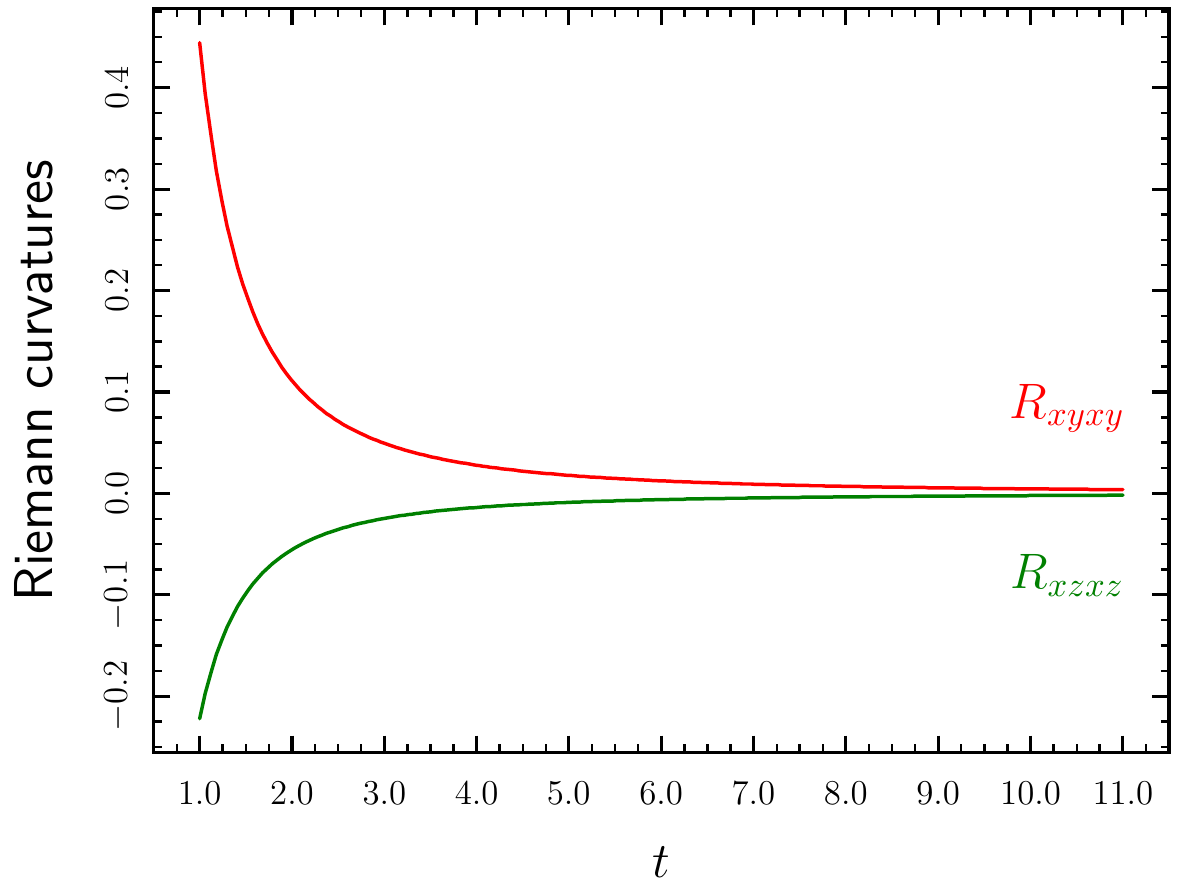}{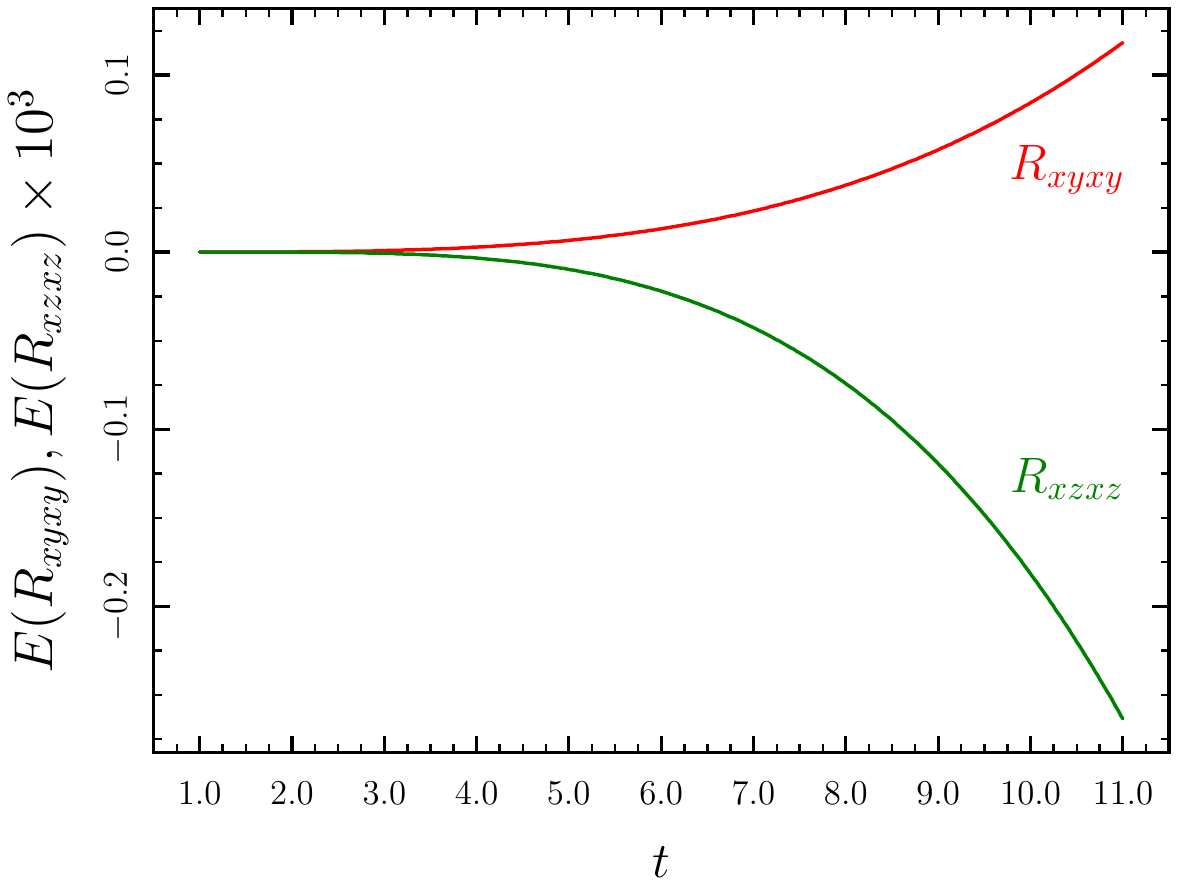}
\caption{\normalfont%
Smooth lattice evolution of $\Rxyxy$ and $\Rxzxz$ (left) and their corresponding
fractional errors (right).}
\label{fig:slgrRabcd}
\end{figure}

The smooth lattice method computes not just the leg-lengths but also the extrinsic and
Riemann curvatures. These can be compared with the exact values
(\ref{eqn:RNCKasner1},\ref{eqn:RNCKasner2}). This leads to the results shown in figures
(\ref{fig:slgrKab}) and (\ref{fig:slgrRabcd}). This again shows that the method tracks the
exact solution very well. Plots such as these are not so easily constructed for the Regge
calculus.

It should also be noted that where Gentle and Miller noticed high frequency oscillations in
some of their data no such oscillations were observed in the Regge solutions described here.
This difference is mostly likely due to the different ways in which the initial data were
set.

The results presented above are all concerned with accuracy and convergence. But equally
important is the computational cost. The $N_x=N_y=N_z=8$ models place very little demand on
memory so the computational cost is dominated by the cpu time. It was found that the Regge
calculus was around \CpuRatio\ times slower than the smooth lattice method. This poor
performance is most likely due two keys elements of the Regge method -- at each time step it
has to solve 14 non-linear equations for 14 leg-lengths at each vertex while also frequently
computing inverse trigonometric and hyperbolic functions. Despite using the best
computational methods available \cite{brewin:2010-01} this gap between the Regge calculus
and the smooth lattice method remained.

\begin{figure}[ht]
\FigPair{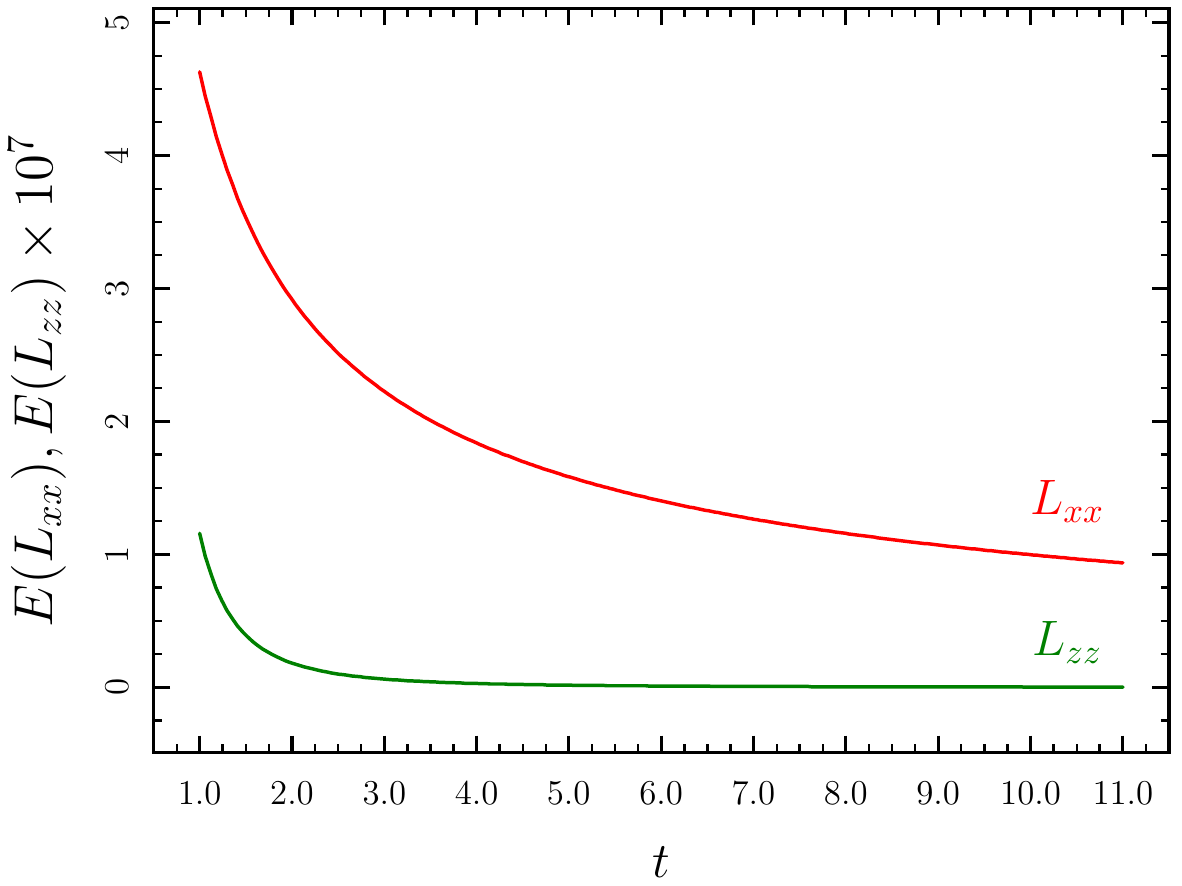}
        {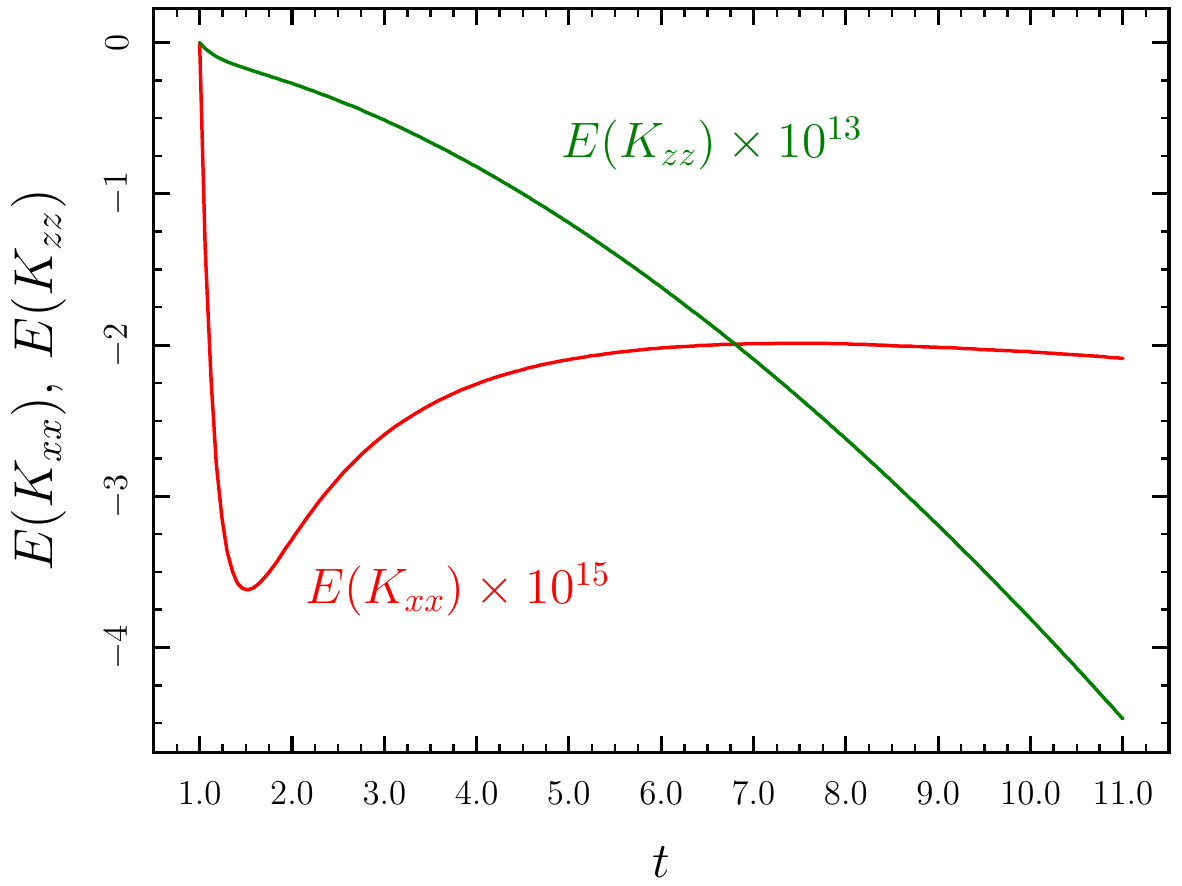}
\vspace{5pt}
\FigPair{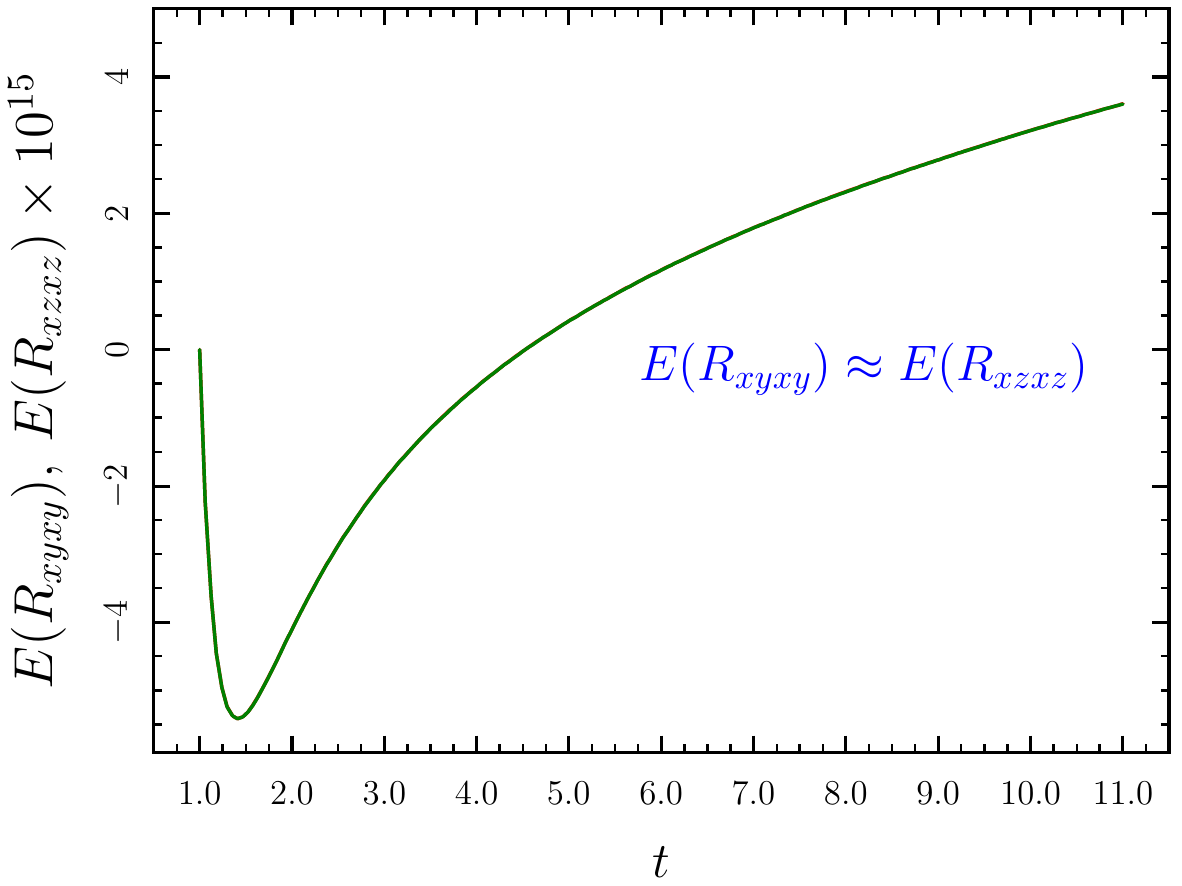}
        {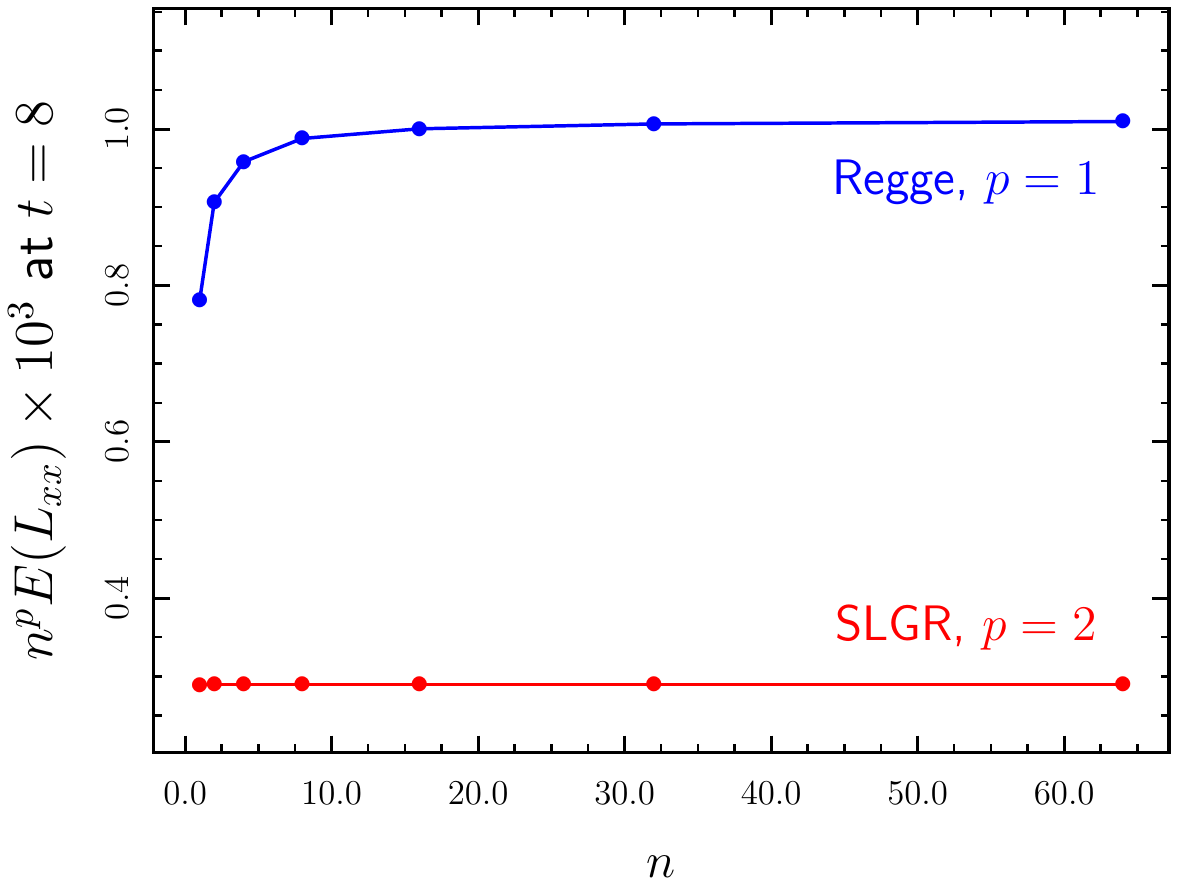}
\caption{\normalfont%
A sample of results for the second evolution scheme for the smooth lattice method. These
results are a considerable improvement over the results from the first evolution scheme. The
Hamiltonian is not plotted here as in this scheme it is zero throughout the evolution.}
\label{fig:slgrScheme2}
\end{figure}

The results from the second evolution scheme for the smooth lattice were found to be
significantly better than for the first scheme with a selection of results shown in figure
(\ref{fig:slgrScheme2}). This scheme also runs much faster than the first scheme (it is
quicker to evolve the coordinates than it is to solve the coupled quadratic equations
(\ref{eqn:RNCLij})). In this case the smooth lattice method is approximately \CpuRatioFast\
times faster than the Regge calculus.

Given that this second scheme works so well the question must be -- why bother with the
first scheme? The simple answer is that the homogeneity of this space-time might be giving
the second scheme advantages not shared with the first scheme. Another cause for concern
with the second scheme is that it weakens the coupling between neighbouring frames. In the
first scheme the evolution of the leg lengths was shared between pairs of frames (one from
each vertex of the leg). In this second scheme the leg lengths are derived from the evolved
coordinates of a single frame. It is not clear what impact this might have on the evolution
of less symmetric initial data. Both schemes should be tested on space-times devoid of any
symmetry before a decision is made on which is to be preferred.

\section{Discussion}
\label{sec:Discuss}

What objective conclusions can be drawn from the results just presented? Arguments may be
made in favour of the formulation of both the Regge calculus and the smooth lattice method
but such arguments are, to a degree, subjective being based on personal preferences. For
example, the simplicity and elegance of the Regge calculus could be given as an argument in
support of the Regge calculus. Equally, a case could be made that the smooth lattice method
is better equipped to deal with differential operators on a lattice than the Regge calculus.
But the real crunch comes when asking which method gives the better numerical results. That
is, given the same initial data on the same lattice, how do the evolved data compare for
accuracy, stability and cost in memory and cpu time? The question of accuracy can not be
properly dealt with here for the simple reason that the initial data were not constructed as
solutions of the respective constraint equations. However, as noted in section
(\ref{sec:GeomData}), the manner in which the initial data is constructed should have no
bearing on the cpu time required to evolve the initial data. Thus the observation that the
Regge calculus is \CpuRatio\ times slower (or worse) than the smooth lattice method must be
taken seriously. If this number can not be reduced then it is hard to see how the Regge
calculus could compete against the smooth lattice method. One approach might be to take
advantage of the massive parallelism available in the Sorkin algorithm. However this is
unlikely to do the trick as the same is true of the smooth lattice method. The questions of
stability and memory are also of little value in this instance as both methods were seen to
be stable over $1<t<11$ and required similar storage for these $8\times8\times8$ lattices.
The upshot is that from the results presented above there is only one meaningful measure,
the total cpu time, and that places the smooth lattice method well ahead of the Regge
calculus.

\appendix

\section{The SLGR source terms}
\label{app:SourceTerms}

In section (\ref{sec:SLGRSourceTerms}) it was noted that an essential step in estimating the
sources terms, such as $\dRyzyzy$ in (\ref{eqn:SLGRdotRabcd7}), requires data to be imported
from one cell to another. The purpose of this section is to provide full details of that
procedure. The following section will apply these ideas to the particular case of the
bi-cubic lattice.

As a simple example, consider the case where the components of a vector $v^\alpha$, defined
at vertex $q$, are to be imported from $\barq$ to $\barp$ (i.e., a transformation between
frames). Since this map occurs in the tangent space of the vertex $q$ the transformation
must be of the form
\begin{align}
	v^\alpha_{q\barp} &= M^\alpha{}_\beta(p,q) v^\beta_{q\barq}
\label{eqn:Import}
\end{align}
for some yet to be determined $4\times4$ matrix $M(p,q)$. The entries in $M(p,q)$ are not
entirely arbitrary as the transformation must preserve scalar products (the transformation
is simply a change of frame at a point). Thus the matrix $M$ is a proper Lorentz
transformation characterised by six parameters -- three boosts and three rotations.

So how might $M(p,q)$ be computed? If four linearly independent vectors at $q$ could be
constructed, say $v^\alpha_i,\>i=1,2,3,4$, and if their components in each frame could be
found (by means other than from the above equation) then a system of equations such as
\begin{align}
	v^\alpha_{iq\barp} &= M^\alpha{}_\beta(p,q) v^\beta_{iq\barq}\quad\quad i=1,2,3,4
\label{eqn:ImportSystem}
\end{align}
could be proposed for the unknown matrix $M(p,q)$. The assumption that the four vectors at
$q$ are linearly independent ensures that a unique solution for $M(p,q)$ exists. This matrix
can then be used to import data of any kind from $\barq$ to $\barp$, for example
\begin{align}
	K^{\alpha\beta}_{q\barp} = M^\alpha{}_\mu(p,q)M^\beta{}_\nu(p,q)K^{\mu\nu}_{q\barq}
\label{eqn:ImportExample}
\end{align}
The challenge now is to first identify four linearly independent vectors
$v^\alpha_i,\>i=1,2,3,4$ and second their components in each frame. One of the vectors,
$v^\alpha_1$, can be taken to be the future pointing time-like normal $n^\alpha$ to the
Cauchy surface at $q$. But what of the remaining three vectors? This is where the legs of
the lattice enters the picture -- they provide the necessary information by which the
vectors and their components can be constructed. Thus the remaining vectors,
$v^\alpha_i,\>i=2,3,4$, are chosen to be unit tangent vectors to three of the legs attached
to $q$. There are of course many legs attached to $q$, so which three should be chosen?
Since the unit vectors to these legs are about to be used to compute $M(p,q)$ it makes sense
to choose legs that are shared by the cells $p$ and $q$. Now choose one of the three legs to
be the leg joining $p$ to $q$. The remaining two legs can be freely chosen. This
construction guarantees the linear independence of the four vectors at $q$ with one possible
exception -- when the last pair of legs just described coincide. This degenerate case is
extremely unlikely to occur in practice\footnote{This will occur only when the transition
functions from $\barq$ to $\barp$ fail to be invertible, equivalently, the 3-volume shared
by $\barq$ and $\barp$ vanishes.} and thus will be ignored.

The Riemann normal coordinates for any cell can be computed according to the procedure
described in \cite{brewin:2010-03}. This in turn allows the tangent vectors to each leg to
be computed at $q$ in each frame. In particular, the components of the unit tangent vector,
at $q$, to the leg that joins $q$ to $r$ are given by
\begin{align}
	v^\alpha_{q\barq}(q,r) L_{qr} &= x^\alpha_{r\barq} - x^\alpha_{q\barq}\label{eqn:TangentVec1}\\
	v^\alpha_{q\barp}(q,r) L_{qr} &= x^\alpha_{r\barp} - x^\alpha_{q\barp} + \BigO{R L^3}\label{eqn:TangentVec2}
\end{align}
where $x^\alpha$ are the Riemann normal coordinates. The first equation follows directly
from the definition of Riemann normal coordinates while the second is the leading order
expansion of the solution to the geodesic boundary value problem for the geodesic that
connects $q$ to $r$. For full details see \cite{brewin:2009-03}. Note that although these
vectors are tangent to the legs of the lattice they will not in general be orthogonal to the
normal $n^\alpha$. This follows form the simple observation that the geodesic that joins a
pair of vertices need not lie within the Cauchy surface but will in general be a chord for
that pair of vertices. However, since the time coordinate of each vertex satisfies
$2t = -K_{\alpha\beta}x^\alpha x^\beta$ (see \cite{brewin:2010-03}) it follows that
$n^\alpha v_\alpha = \BigO{L^2}$.

Consider now the time-like vector normal $n^\alpha$. In the local Riemann normal coordinates
with zero shift and unit lapse, the components for $n^\alpha$ in each frame are simply
\begin{align*}
	n^\alpha_{p\barp} = n^\alpha_{q\barq} = \delta^\alpha_t
\end{align*}
while the values for $n^\alpha_{q\barp}$ can be estimated by a local Taylor series around
$p$
\begin{align*}
	n^\alpha_{q\barp} &= n^\alpha_{p\barp} + n^\alpha{}_{;\beta}\Dx^\beta (p,q) + \BigO{L^2}
\end{align*}
However, for a unit lapse function,
\begin{align*}
	n_{\mu;\nu} = - K_{\mu\nu}
\end{align*}
where $K^\alpha{}_\beta$ are the components of the extrinsic curvature at $p$ in the frame
$\barp$. Thus the above equation can be re-written as\footnote{The same result is obtained
for the case where the lapse is a smooth function across the lattice.}
\begin{align}
	n^\alpha_{q\barp} = \left( \delta^\alpha{}_\beta
	                          + n_\beta K^\alpha{}_\gamma x^\gamma_{q\barp}
							  \right) n^\beta_{q\barq}
	                   + \BigO{L^2}
\label{eqn:ImportNormal}
\end{align}
This completes the construction of all four vectors in each frame and thus the
transformation matrix $M^\alpha{}_\beta(p,q)$ can be computed from (\ref{eqn:ImportSystem}).

\section{The bi-cubic lattice}
\label{app:BiCubic}

Consider for the moment the broad picture presented in section (\ref{sec:SLGR}) in which a
smooth lattice is constructed from a known smooth geometry $S$. Choose any set of
coordinates over some subset $V$ of $S$. The intersection of the various coordinate planes
in $V$ produces a set of cubic cells that can naturally be interpreted as the cells of a
lattice. In this construction the unit vectors tangent to the legs of the cells will vary
smoothly across the lattice. It is also clear, by increasing the density of the coordinate
planes in $V$, that the unit vectors in one cell will converge to the corresponding unit
vectors in a neighbouring cell. Thus the matrix $M(p,q)$ for a pair of cells, $p$ and $q$,
not only reduces to the identity in the limit as $p$ approaches $q$ but it should do so
smoothly. Consequently, for a bi-cubic lattice, $M(p,q)$ should be of the form
\begin{align}
	M^\alpha{}_\beta(p,q) = \delta^\alpha{}_\beta + m^\alpha{}_\beta(p,q) + \BigO{L^2}
\label{eqn:ImportExpand}
\end{align}
where $m(p,q)$ is another $4\times4$ matrix at $q$ with entries $\BigO{L}$ and where $L$ is
a typical length scale for the cell (e.g., the length of the leg joining $p$ to $q$).

Note that this result does not necessarily apply to other lattices, such as a simplicial
lattice \footnote{However, the smooth variation of the time like normal to the Cauchy
surface would allow $M(p,q)$ to be factored in the form $M(p,q) = (I+B(p,q))R(p,q)$ where
$I$ is the identity matrix, $I + B(p,q)$ is a boost matrix with $B(p,q) = \BigO{L}$ and
$R(p,q)= \BigO{1}$ is a pure spatial rotation matrix.}.

The matrix $m(q,p)$, like $M(p,q)$, is subject to the constraint that the transformations
must preserve scalar products. Noting that the metric in each Riemann normal frame is of the
form $\diag(-1,1,1,1)+\BigO{RL^2}$ it is easy to see that this leads to the following
constraint on $m(p,q)$
\begin{align}
	0 = g_{\alpha\rho} m^\rho{}_\beta(p,q)
	  + g_{\beta\rho}  m^\rho{}_\alpha(p,q)
\label{eqn:SkewSymm1}
\end{align}
That is, the $m_{\mu\nu}$ define a skew-symmetric $4\times4$ matrix determined by just six
independent entries (corresponding to the three boosts and three rotations).

Note that equation (\ref{eqn:ImportNormal}) is already in the form (\ref{eqn:Import}) and
thus it provides some immediate information about $m(p,q)$. It follows from
(\ref{eqn:ImportExpand},\ref{eqn:SkewSymm1}) and (\ref{eqn:ImportNormal}) that $m(p,q)$ is
of the form
\begin{align}
	m^\alpha{}_\beta(p,q)
	           = n_\beta K^\alpha{}_\gamma x^\gamma_{q\barp}
				  - n^\alpha K_{\beta\gamma} x^\gamma_{q\barp}
				  + s^\alpha{}_\beta(p,q) + \BigO{L^2}
\label{eqn:SkewSymm2}
\end{align}
where $s(p,q)$ is another $4\times4$ matrix subject to
\begin{align}
	0 &= s^\alpha{}_\beta(p,q) n^\beta{}_{q\barq}\label{eqn:ZeroNormal1}\\
	0 &= s^\alpha{}_\beta(p,q) n_{\alpha q\barq}\label{eqn:ZeroNormal2}
\end{align}
(i.e., the matrix $s(p,q)$ has no normal component, it is a purely spatial matrix). In the
adapted Riemann normal coordinates where $n^\mu = \delta^\mu_t$ and
$g_{\mu\nu}=\diag(-1,1,1,1)$ this requires the first row and column of $s(p,q)$ to be zero.
Furthermore, the constraints (\ref{eqn:SkewSymm1}) shows that the remaining $3\times3$ sub
matrix of $s(p,q)$ must be skew symmetric. Thus $s(p,q)$ describes the three rotations while
the remaining terms in $m(p,q)$ describe the three boosts.

The remaining entries in $s(p,q)$ can now be obtained by applying the transformation
(\ref{eqn:ImportSystem}) to the three vectors $v^\mu_i,\>i=2,3,4$. This leads to a
$3\times3$ system for $s(p,q)$
\begin{align}
	v^\alpha_{iq\barp}
 - v^\alpha_{iq\barq}
 = s^\alpha{}_\beta(p,q) v^\beta_{iq\barq}\quad\quad i=2,3,4
\label{eqn:3x3sys}
\end{align}
where the Greek indices are now restricted to cover the $3\times3$ sub-matrix of $s(p,q)$.
This is an overdetermined system of equations for the three non-zero entries of
$s^\alpha{}_\beta(p,q)$.

In appendix (\ref{app:SourceTerms}) it was argued that the three vectors
$v^\mu_{iq},\>i=2,3,4$ could be chosen as the unit vectors tangent to the legs at $q$. But
equally any (invertible) linear combination of these vectors could also be used. This
freedom can be used, as described below, to produce a near diagonal $3\times3$ system of
equations for $s(p,q)$.

\begin{figure}[t]
\FigPair{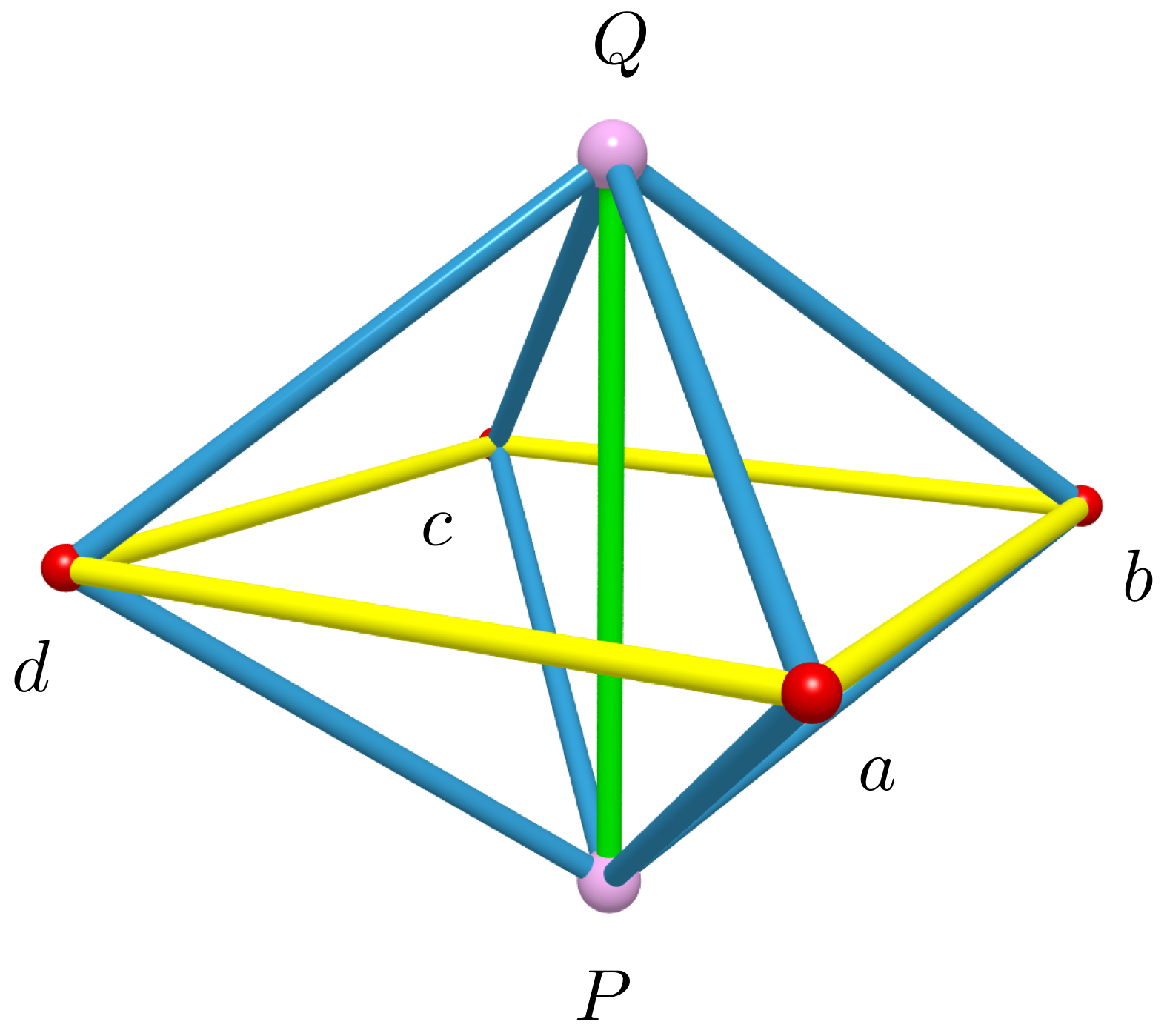}{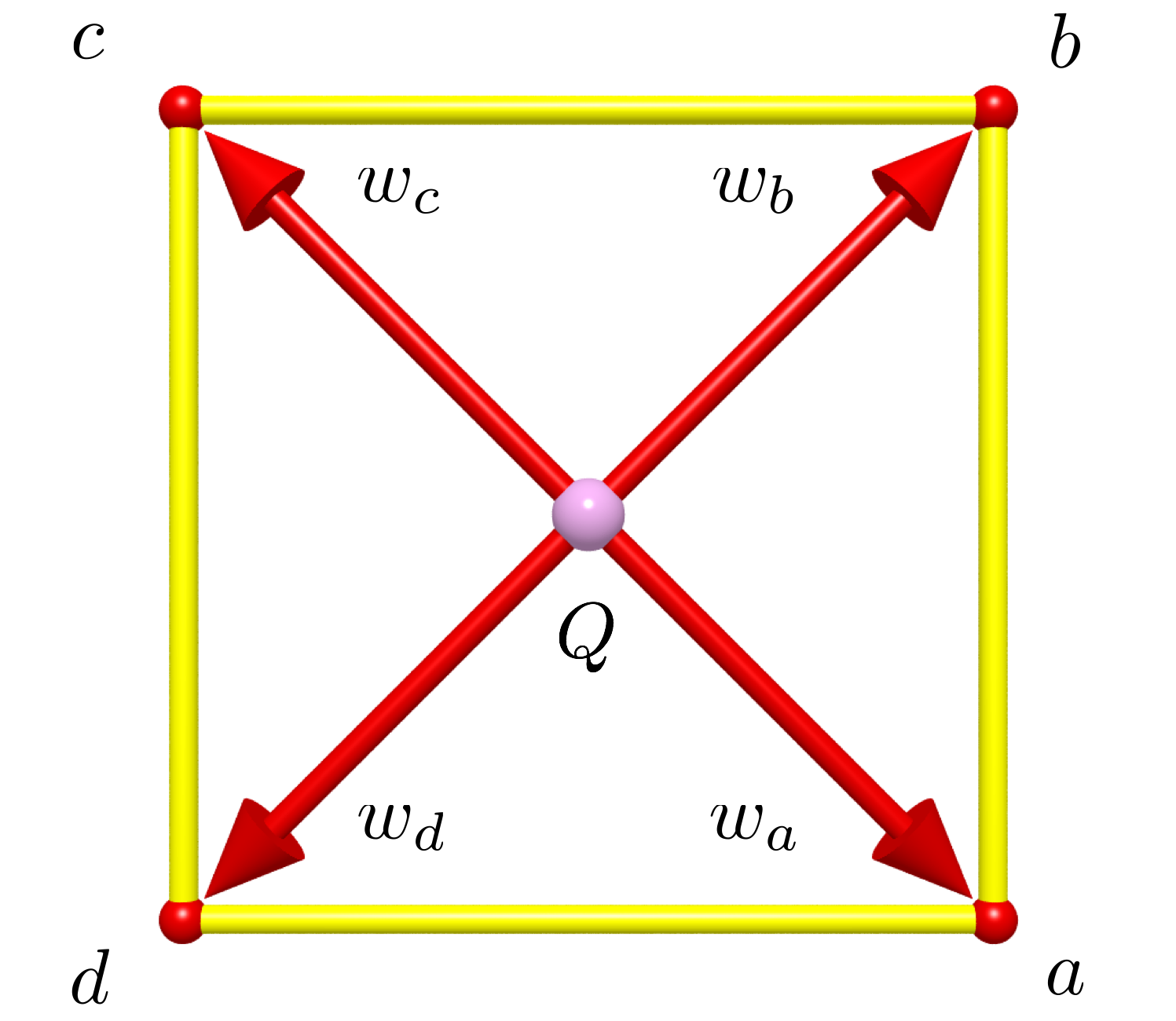}
\caption{\normalfont%
The set of legs shared by a pair of cells (left) and, viewed from above looking down on
vertex $Q$, the set of vectors (right) used when constructing the map used to import the
data from one frame to the other. Note that the labels $w_a,w_b\cdots$ are abbreviations for
the vectors $w(a,q),w(b,q)\cdots$ used in the text.}
\label{fig:BiCubicShared}
\end{figure}

The typical set of legs shared by a pair of cells in a bi-cubic lattice are shown in figure
(\ref{fig:BiCubicShared}). This consists of five legs attached to $q$ and correspondingly,
five tangent vectors $w^\alpha(i,q),\>i=p,a,b,c,d$. However the discussion above requires a
selection of three linearly independent vectors at $q$. Many choices are possible, such as
\begin{align}
	v^\alpha_2 &= \lambda_2 w^\alpha(p,q)\label{eqn:ThreeVec1}\\
	v^\alpha_3 &= \lambda_3 \left(w^\alpha(a,q)-w^\alpha(b,q)-w^\alpha(c,q)+w^\alpha(d,q)\right)\label{eqn:ThreeVec2}\\
	v^\alpha_4 &= \lambda_4 \left(w^\alpha(a,q)+w^\alpha(b,q)-w^\alpha(c,q)-w^\alpha(d,q)\right)\label{eqn:ThreeVec3}
\end{align}
with $\lambda_i,\>i=2,3,4$ chosen so that each $v_i,\>i=2,3,4$ is a unit vector. The
justification for this choice is that, in the case of a flat lattice, the three vectors are
aligned to the local coordinate axes. This is not an important point and many other choices
might work equally as well. No such variations were tried for this paper. Note the change in
notation -- the $v_i,\>i=2,3,4$ are not tangent to the legs, that role is now played by the
$w(i,q),\>i=p,a,b,c,d$.

As noted above, the equations (\ref{eqn:3x3sys}) are an over determined system -- they
provide nine equations for just three non-zero entries in $s(p,q)$. The system could be
solved using a least squares method but that is expensive (this calculation is at the heart
of the evolution equations so computational cost is an important issue). The direct
approach, as adopted in this paper, is to select just three of the nine equations and solve
the resulting $3\times3$ system using standard matrix methods. This worked very well for
this lattice and this space-time. How were the three equations chosen? Consider for example
the case were the lattice is almost flat. Then as noted above the three vectors
$v^\alpha_i,\>i=2,3,4$ will be closely aligned to the coordinate axes. Thus suppose $v_2$ is
approximately aligned to the $x$-axis, $v_3$ to the $y$-axis and $v_4$ to the $z$-axis. Thus
$v^x_2\approx1,\>v^y_3\approx1,\>v^z_4\approx1$ while the remaining components have $\vert
v^\alpha_i\vert = \BigO{L}$. Put
\begin{align}
	s(p,q) = \begin{bmatrix}0&\Sxy&\Sxz\\-\Sxy&0&\Syz\\-\Sxz&-\Syz&0\end{bmatrix}
\label{eqn:ABCmatrix}
\end{align}
and then select the following three of the nine equations (\ref{eqn:3x3sys})
\begin{align}
	\begin{bmatrix}
	   v^y_{2q\barp} - v^y_{2q\barq}\\[3pt]
	   v^x_{4q\barp} - v^x_{4q\barq}\\[3pt]
	   v^z_{3q\barp} - v^z_{3q\barq}\end{bmatrix}
   &=
	\begin{bmatrix}
	  -v^x_{2q\barq}&0&v^z_{2q\barq}\\[3pt]
	   v^y_{4q\barq}&v^z_{4q\barq}&0\\[3pt]
	   0&-v^x_{3q\barq}&-v^y_{3q\barq}\end{bmatrix}
	\begin{bmatrix}\Sxy\\[3pt]\Sxz\\[3pt]\Syz\end{bmatrix}
\label{eqn:Sabeqtns}
\end{align}
By inspection it is easy to see that the diagonal entries of the coefficient matrix are
close to $\pm1$ while the remaining entries are small. Thus this $3\times3$ equation is
non-singular and easily solved for $\Sxy,\Sxz$ and $\Syz$.

Even though the above equations (\ref{eqn:Sabeqtns}) were selected on the assumption that
the lattice was almost flat they can be expected, on continuity arguments, to be useable in
cases where the lattice is not approximately flat. This is consistent with the results in
section (\ref{sec:Results}) -- at no point in the evolution did the $3\times3$ set of
equations have a condition number not close to one.

Once the matrices $m(p,q)$ are known for each of the vertices $q$ that surround $p$ then the
derivatives of any tensor can be estimated at $p$ using a finite difference method. Consider
the simple case where the spatial derivatives $u^\mu_{,x}$, $u^\mu_{,y}$ and $u^\mu_{,z}$ of
some vector $u^\mu$ are required at $p$. Begin by writing out a short Taylor series
\begin{align}
	u^\mu_{q\barp} = u^\mu_{p\barp} + u^\mu_{,\nu p\barp}x^\nu_{q\barp} + \BigO{L^2}
\end{align}
then use equations (\ref{eqn:Import}) and (\ref{eqn:ImportExpand}) to obtain
\begin{align}
	u^\mu_{,\nu p\barp}x^\nu_{q\barp} = u^\mu_{q\barq} - u^\mu_{p\barp} + m^\mu{}_\alpha (p,q) u^\alpha_{q\barq}
\end{align}
Note that the time coordinate $x^t_q = \BigO{L^2}$ and thus the left hand side only contains
the three spatial derivatives. This system of equations can be written down, to $\BigO{L}$,
for each of the vertices $q$ and once again leads to an overdetermined system for the
spatial derivatives. For the bi-cubic lattice there is a rather simple reduction to a well
defined $3\times3$ system of equations. Define $\Delta u^\mu(p,q)$ by
\begin{align}
	\Delta u^\mu (p,q) = u^\mu_{q\barq} - u^\mu_{p\barp} + m^\mu{}_\alpha (p,q) u^\alpha_{q\barq}
\end{align}
Then build the following set of equations
\begin{align}
u^\mu_{,\nu p\barp}\left(x^\nu_{10}-x^\nu_{12}\right) &= \Delta u^\mu(0,10) - \Delta u^\mu(0,12)\\
u^\mu_{,\nu p\barp}\left(x^\nu_{11}-x^\nu_{9}\right)  &= \Delta u^\mu(0,11) - \Delta u^\mu(0,9)\\
u^\mu_{,\nu p\barp}\left(x^\nu_{13}-x^\nu_{14}\right) &= \Delta u^\mu(0,13) - \Delta u^\mu(0,14)
\end{align}
in which the integers 10,12, etc. are the vertex labels (see figure
(\ref{fig:BiCubicRNCs})). This provides, for each choice of $\mu$, a $3\times3$ system of
equations for $u^\mu_{,\nu p\barp}$. The same ideas can be used to compute the $R_{xyxy,z}$
etc.

\section{Coordinates in the bi-cubic cell}
\label{app:BiCubicRNCs}

In \cite{brewin:2010-03} a general procedure was given for computing the coordinates
$x^\mu_i$ for each vertex in a cell. For the bi-cubic lattice the particular details are as
follows.

\begin{figure}[ht]
\Figure{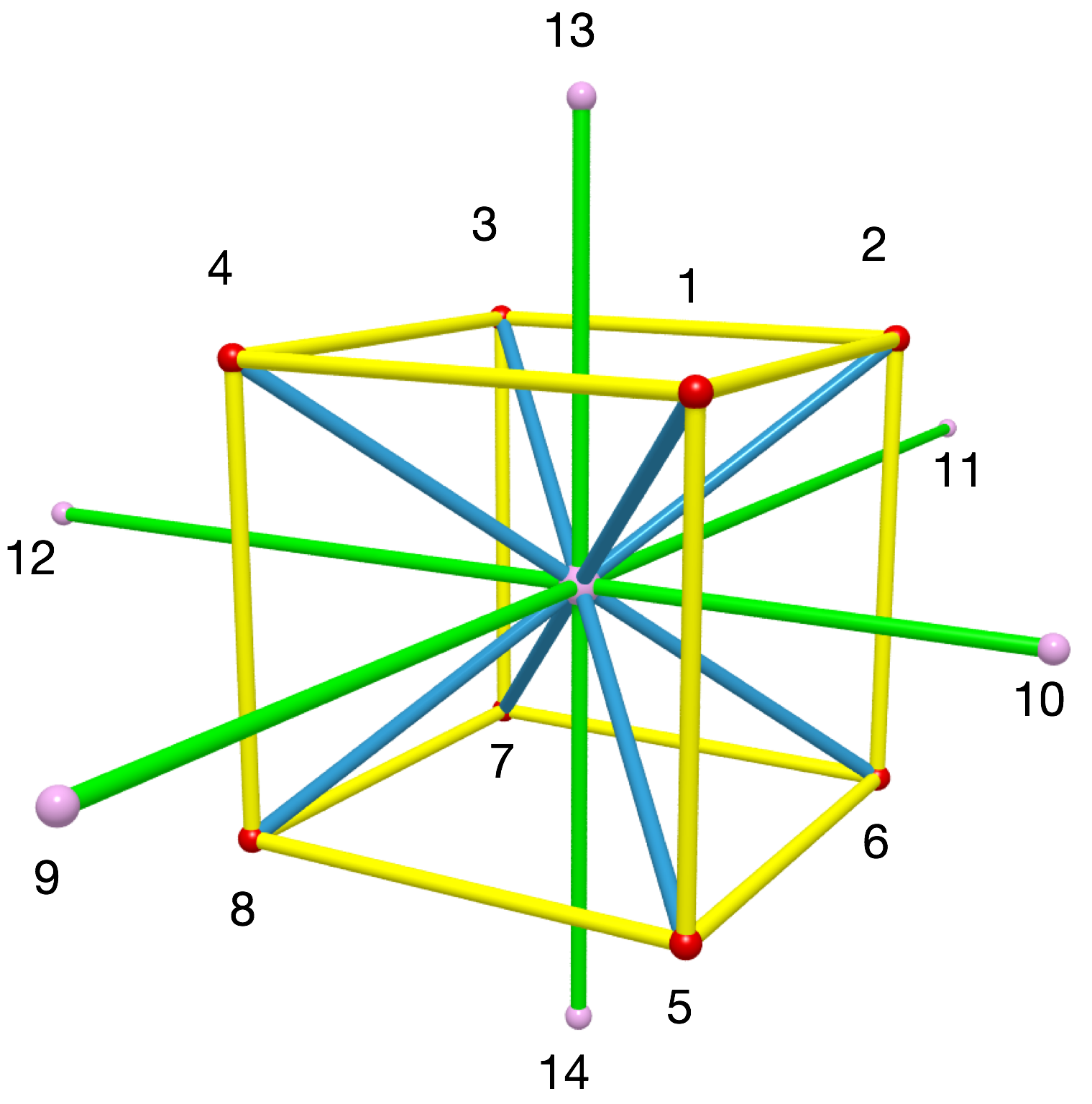}
\caption{\normalfont%
The bi-cubic cell with labels assigned to each vertex. The label for the central vertex is 0
but is excluded from this figure to avoid clutter. The local Riemann normal coordinates
$(x,y,z)$ are chosen so that vertex 0 has coordinates $(0,0,0)$, vertex 13 lies on the
positive $z$-axis and vertex 10 lies in the $xz$-plane with $x_{10}>0$.}
\label{fig:BiCubicRNCs}
\end{figure}

The fifteen vertices of the cell are numbered for 0 to 14 as per figure
(\ref{fig:BiCubicRNCs}). Consider three vertices $i$, $j$ and $k$ that form a tetrahedron
attached to the central vertex. Suppose also that the coordinates for vertices $i$ and $j$
have been computed. Then the coordinates $x^\mu_k$ for vertex $k$ can be computed using
\def\Lsq{L^2}
\def\Lsqoi{\Lsq_{oi}}
\def\Lsqoj{\Lsq_{oj}}
\def\Lsqok{\Lsq_{ok}}
\def\Lsqij{\Lsq_{ij}}
\def\Lsqik{\Lsq_{ik}}
\def\Lsqjk{\Lsq_{jk}}
\begin{equation*}
x^\mu_k = P x^\mu_i + Q x^\mu_j + R n^\mu_{ij}
\end{equation*}
where
\begin{gather*}
P=\frac{m_{ik}\Lsqoj-m_{jk}m_{ij}}{\Lsq_n}\qquad
Q=\frac{m_{jk}\Lsqoi-m_{ik}m_{ij}}{\Lsq_n}\\[5pt]
R=\frac{\left(\Lsqok-P^2\Lsqoi-Q^2\Lsqoj-2PQm_{ij}\right)^{1/2}}{L_n}\\[5pt]
n^\mu_{ij} = g^{\mu\nu} \epsilon^{xyz}_{\nu\alpha\beta} x^\alpha_i x^\beta_j\qquad
\Lsq_n = \Lsqoi\Lsqoj-m^2_{ij}
\end{gather*}
and where the $m_{ab}$ are defined by
\begin{gather*}
2m_{ij} = \Lsqoi+\Lsqoj-\Lsqij\\[5pt]
2m_{ik} = \Lsqoi+\Lsqok-\Lsqik\quad\quad\quad
2m_{jk} = \Lsqoj+\Lsqok-\Lsqjk
\end{gather*}
Note the $n^\mu_{ij}$ is a vector normal to the triangle $(oij)$ pointing towards vertex $k$
(this can always be achieved by swapping $i$ and $j$ if required).

The above procedure can be applied to all of the vertices provided initial values have been
set for the first pair. This last step amounts to fixing the rotational freedoms. The frame
chosen for this paper has the vertex 13 lying on the positive $z$-axis, i.e., $x_{13} =
(0,0,x^z_{13})$ and vertex 10 lying somewhere in the $xz$-plane, i.e., $x_{10} =
(x^x_{10},0,x^z_{10})$. However the cell does not contain the triangle $(0,10,13)$ and thus
the coordinates for vertex 10, in this gauge, are not immediately available. This problem
can be overcome by first choosing an intermediate gauge where vertex 1 lies in the
$xz$-plane. In this gauge the coordinates for vertices 13 and 1 are easily computed,
\begin{align}
	x_{13} = (0,0,L_{0,13}),\qquad
	x_1 = ((L^2_{0,1}-m^2/L^2_{0,13})^{1/2},0,m/L_{0,13})
\label{eqn:GaugeCond}
\end{align}
with $m=(L^2_{0,1}+L^2_{0,13}-L^2_{1,13})/2$. This allows the coordinates for the remaining
13 vertices to be computed. Finally a rotation in the $xy$-plane is applied to cell to set
$x^y_{10}=0$ and $x^x_{10}>0$.

Table (\ref{tbl:BiCubicRNCs}) shows the order in which the vertex coordinates were computed
(in the intermediate gauge). The notation $(k;i,j)$ indicates that the coordinates for
vertex $k$ are computed form the known coordinates for vertices $i$ and $j$. Note that the
order of $i$ and $j$ is important, they must be chosen so that the normal vector $n_{ij}$
points towards vertex $k$.
\begin{table}[t]
	\centerline{
	\begin{tabular}{cccccc}
		\hline
		\vrule height 14pt depth 0pt width 0pt
		(2;13,1)&(3;13,2)&(4;13,3)&(9;1,4)&(10;,2,1)&(11;3,2)\\
		(12;4,3)&(5;1,9)&(6;2,10)&(7;3,11)&(8;4,12)&(14;6,5)
		\vrule height 12pt depth 8pt width 0pt\\
		\hline
	\end{tabular}}
   \caption{The order in which the vertices are computed in the intermediate gauge. The
   entries should be read from left to right and top to bottom. Each entry is of the form
   (k;i,j) and this indicates that the coordinates of vertex $k$ are computed from the known
   coordinates of vertices $i$ and $j$. The coordinates of the central vertex are $(0,0,0)$
   while the coordinates for vertices 13 and 1 are given by equation (\ref{eqn:GaugeCond}).}
	\label{tbl:BiCubicRNCs}
\end{table}

\section{Evolution of the RNC's}
\label{app:EvolveRNCs}

In the standard Cauchy IVP there are two world-lines through any event, one is the observer's
world-line the other is the integral curve of the normal vector to the Cauchy surface
containing that event. In a lattice method allowance must be made for a possible third
world-line, the world-line of a vertex. Just as the observer's world-line can be freely chosen
in a given space-time so too can the vertex world-line. In the absence of any preferred
spatial directions (e.g., the velocity vector of a fluid flow) a simple choice is to require
each vertex to follow the integral curve of the normal vector and that in every cell the
central vertex is forever located at the spatial origin (i.e., the shift vector vanishes at
each central vertex). This is the choice made in this paper.

Let $x^\mu_{q\bar p}(t)$ be the coordinates along the world-line of vertex $q$ in the frame
of $p$. Since this world-line coincides with the integral curve of the normal at $q$ it
follows that the coordinates evolve according to
\begin{align}
	\frac{dx^\mu_{q\bar p}}{dt} = n^\mu_{q\bar p}
\end{align}
which upon using (\ref{eqn:ImportNormal}) and $n^\mu_{q\bar q}=\delta^\mu_t$ leads to
\begin{align}
   \frac{dx^\mu_{q\bar p}}{dt} = -K^\mu{}_\nu x^\nu_{q\bar p} + \BigO{L^2}\qquad\hbox{for }\mu=x,y,z
\end{align}
These equations can be used to evolve the spatial part of the Riemann normal coordinates of
each vertex within a cell. At any time the leg lengths can be recovered using
(\ref{eqn:RNCLij}).

\clearpage


\input{./paper.bbl}           

\end{document}

%% file: paper.bbl
\providecommand{\href}[2]{#2}\begingroup\raggedright\endgroup